\documentclass[pdflatex,sn-aps]{sn-jnl}

\usepackage{hyperref}
\usepackage{graphicx}%
\usepackage{multirow}%
\usepackage{amsmath,amssymb,amsfonts}%
\usepackage{amsthm}%
\usepackage{mathrsfs}%
\usepackage[title]{appendix}%
\usepackage{xcolor}%
\usepackage{textcomp}%
\usepackage{manyfoot}%
\usepackage{booktabs}%
\usepackage{algorithm}%
\usepackage{algorithmicx}%
\usepackage{algpseudocode}%
\usepackage{listings}%
\usepackage{soul} 
\usepackage{verbatim} 
\usepackage{caption}
\usepackage[dvipsnames]{xcolor} 
\usepackage[T1]{fontenc} 
\usepackage{booktabs} 
\usepackage{makecell} 

\usepackage{caption}

\theoremstyle{thmstyleone}%
\theoremstyle{thmstyletwo}%

\theoremstyle{thmstylethree}%
\newcommand{\ar}{\textcolor{black}} 
\newcommand{\arr}{\textcolor{black}} %

\begin{document}

\title[Spherical Balls Settling Through a Quiescent
Cement Paste]{Spherical Balls Settling Through a Quiescent
Cement Paste Measured by X-ray Tomography: 
Influence of the Paste Thixotropy}

\author[1]{\fnm{Subhransu} \sur{Dhar} \href{https://orcid.org/0000-0003-2484-194X}{\includegraphics[scale=0.035]{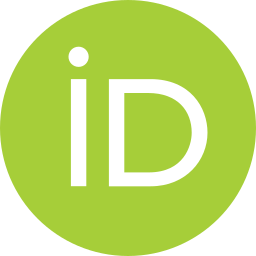}}}\email{subhransu.dhar@tuwien.ac.at} 

\author[2]{\fnm{Eduardo} \sur{Machado-Charry} \href{https://orcid.org/0000-0001-6276-7627}{\includegraphics[scale=0.035]{figs/256px-ORCID_iD.png}}}\email{machadocharry@tugraz.at}

 \author[2]{\fnm{Robert} \sur{Schennach} \href{https://orcid.org/0000-0001-7477-0030}{\includegraphics[scale=0.035]{figs/256px-ORCID_iD.png}}}\email{robert.schennach@tugraz.at}

 \author[1]{\fnm{Teresa} \sur{Liberto} \href{https://orcid.org/0000-0002-4985-9110}{\includegraphics[scale=0.035]{figs/256px-ORCID_iD.png}}}\email{teresa.liberto@tuwien.ac.at}

\author*[1]{\fnm{Agathe} \sur{Robisson} \href{https://orcid.org/0000-0002-2434-6175}{\includegraphics[scale=0.035]{figs/256px-ORCID_iD.png}}}\email{agathe.robisson@tuwien.ac.at}

\affil[1]{\orgdiv{Institute of Material Technology, Building Physics, and Building Ecology}, \orgname{TU~Wien}, \orgaddress{\street{Karlsplatz~13/207}, \postcode{1040},  \city{Wien}, \country{Austria}}}

 \affil[2]{\orgdiv{Institute of Solid State Physics}, \orgname{TU Graz}, \orgaddress{\street{ Petersgasse 16/II}, \postcode{8010}, \city{Graz}, \country{Austria}}}

This article should be referred to as :\\

Dhar, S.,  Machado-Charry, E., Schennach, R., Liberto, T., Robisson, A.. Spherical balls settling through a quiescent cement paste measured by X-ray tomography: Influence of the paste thixotropy. \textit{Journal of Non-Newtonian Fluid Mechanics} 351 105678 (2026)
https://doi.org/10.1016/j.jnnfm.2026.105678

\abstract{
The settling of spherical balls in quiescent fresh
cement pastes of increasing age at rest is studied. Metallic spheres with radii of 2, 2.5 and 3\,mm are dropped into the paste and allowed to settle, while their position is tracked using X-ray tomography. 
The instantaneous velocity of the spheres, calculated from their movement, is observed to be quasi-constant during their fall, and an average is estimated.
The results show that the average velocity of the balls decreases logarithmically with paste age at rest until ball stoppage, for all three ball sizes.
In parallel, the rheological properties of the cement paste are measured using a rheometer equipped with a vane geometry. The evolution of the paste static yield stress over time is evaluated, and proves to be a reliable predictor for ball stoppage.
Finally, thixotropic models of increasing complexity are evaluated. These models consider four forms of structural growth and breakdown parameters, and their ability to capture the ball settling velocity as a function of paste age at rest is compared. This emphasizes the importance of considering paste breakdown in relation to shearing of the paste when the ball passes through it.}

\maketitle
\section{Introduction}

Concrete remains an essential material for modern construction, due to its versatility, rapidly developing mechanical properties, and low cost coming from an optimized manufacturing process~\cite{Ashby2016}. The binder holds together large particles such as sand and pebbles, and is typically a blend of ordinary Portland cement (OPC) and other more or less reactive particles~\cite{EN197-1_2023}, \ar{added to reduce the hydration heat, the cost, or the environmental footprint of cement}.
The hydration reaction that transforms the fresh concrete (a fluid) into a strong solid comes from a dissolution and hydration/precipitation mechanism~\cite{bullard2011mechanisms,inbook,ioannidou2016crucial}. 
The binder, in its fresh state (before hardening), exhibits a complex rheological behavior, including thixotropy, viscoelasticity, aging, and shear-thinning or shear-thickening properties~\cite{banfill1990rheology}. 
These rheological properties have to be controlled, especially in technological placements such as additive manufacturing~\cite{roussel2006thixotropy,coussot2014yield,liu20233d,el2020critical}. 

\ar{The property of ``thixotropy'' refers to fluids that exhibit an apparent decrease in viscosity upon flow, e.g.,~under a constant shear rate, and a subsequent recovery of the viscosity with time when flow is removed~\cite{tropea2007springer,armstrong2016dynamic}. This behavior is typically associated with the fluid changing microstructure. The latter is disrupted by shear (a process called structure breakdown~\cite{barnes1997thixotropy,mujumdar2002transient}, rejuvenation~\cite{coussot2006aging}, or liquefaction when the disruption is caused by vibration~\cite{barnes1997thixotropy}), while flow-stop promotes structure build-up (a process also called and described by particle aggregation, flocculation or coagulation depending on the fluid type and community~\cite{mewis2009thixotropy,larson2019review}). These changes are slow enough to be observable, in contrast to shear-thinning fluids~\cite{mewis2009thixotropy}, and are, in theory, fully reversible~\cite{barnes1997thixotropy}. 
Real systems, as opposed to ideal thixotropic systems~\cite{larson2015constitutive}, often exhibit some amount of viscoelasticity~\cite{mewis2009thixotropy} or aging~\cite{bonacci2020contact,derec2003aging}. 
A model system often studied consists of fumed silica particles in oil~\cite{dullaert2005thixotropy,armstrong2016dynamic,wei2018multimode}. Even in this model system, viscoelastic effects were captured either through steps in shear rate or shear stress~\cite{dullaert2005thixotropy}, stress relaxation~\cite{dullaert2005stress}, or evolution of elastic modulus with time at rest~\cite{armstrong2016dynamic} 
and are associated with the deformable floc structures resulting from the interparticle attraction forces. 
The theoretical description of thixotropic fluids has generally been done using a scalar parameter $\lambda$ that expresses the degree of structure and varies between 0 (fully broken-down structure) and 1 (fully build-up structure). This parameter is most often used to describe the viscosity~\cite{cheng1965phenomenological,coussot2002avalanche} and the yield stress~\cite{mewis2009thixotropy}. Its time evolution is also often expressed by two different terms, one related to paste destructuration and one related to the paste restructuration~\cite{ferroir2004motion}, each having a different time scale~\cite{barnes1997thixotropy,armstrong2016dynamic}. Models incorporating further aspects to capture the behavior of real industrial materials also exist, see for example the work of de Souza Mendes on thixotropic elasto-visco-plastic (TEVP) materials~\cite{desouzamendes2012critical}. }

Fresh cement paste, and by extension fresh concrete, have been described as thixotropic fluids~\cite{banfill1990rheology,roussel2006thixotropy,wallevik2005thixotropic,marchesini2019irreversible}, owing to the strong interparticle forces that develop between the hydrating cement particles~\cite{roussel2006thixotropy,jarny2008modelling} (size $\sim$ $10~\mu  \mathrm{m}$, density $\sim$  $3150~\mathrm{kg/m}^3$, typically dispersed in water at a solid volume fraction $\sim$ 40\%), and lead to  \ar{their aggregation into a 3D-network}~\cite{prabhu2018multi,roussel2011understanding}. At rest, the particle network gets stronger and stiffer, and under shear, it breaks down, giving the cement paste its thixotropic behavior. \ar{In the first hour after mixing cement and water, the behavior of the fresh cement paste can be considered reversible~\cite{roussel2006thixotropy}. In other terms, the paste can be re-sheared after various resting periods, and will recover a very similar evolution of viscosity and yield stress with time. At some point, the hydration reactions accelerate, and irreversible evolutions in rheological properties become dominant, i.e.,~the properties cannot be recovered by intensive mixing (shearing). Such irreversible evolutions are often described by the term ``aging''. At very early age, on the other hand, the paste flocculation is reversible, and the fluid can be described as ``thixotropic''.
Although a complete microstructural picture is still under discussion, a basic picture is drawn here, according to literature~\cite{han2024understanding,michel2024structural} and to our own view~\cite{dhar2024discrepancies,liberto2022small}, that supports the reversibility hypothesis. As soon as cement gets in contact with water, a dissolution/precipitation begins. During this process, nano-particles of calcium-silicate-hydrate (C-S-H) with a size of approximately 5~nm \citep{allen1987development,jonsson2005controlling,ioannidou2016crucial} precipitate at the surface of cement particles and in the interstitial fluid. 
Despite their low concentration at this early age, i.e.,~less than a few percent \citep{scherer2012hydration}, the interactions between C-S-H particles and between C-S-H and charged cement particles, facilitated by an interstitial solution of high ionic strength \citep{bullard2011mechanisms,goyal2021physics}, contribute to create a particle network,
and the paste
develops almost immediately a yield stress that prevents the sedimentation of the dense cement particles. 
The micron-sized cement particle network that builds up through nano-sized hydrate particles, forming bridges, was described indirectly through rheological and microstructural models~\cite{nachbaur2001dynamic,prabhu2018multi}, and visualized by cryo-microscopy in samples a few minutes old~\cite{zingg2008microstructure}.
When subjected to shear, some of the C-S-H bridges break, resulting in the breakup of the multi-scale network of micron-sized cement and nano-sized C-S-H particles (a process we refer to as fluidization) \citep{banfill1991rheology}. This process exhibits macroscopic reversibility over a time frame of about one hour, giving rise to the thixotropic properties of cement \citep{lootens2004gelation,struble1995viscosity,wallevik2005thixotropic,marchesini2019irreversible,link2023mechanisms}. 
After some time, the yield stress is high enough, reaching 20--40~kPa~\cite{zheng2012prediction}, to prevent the penetration of a needle (standard Vicat test~\cite{DINEN1963_2017}), a point often defined as the setting of the cement paste. It typically occurs after a few hours at rest, and is highly dependent upon the presence of additives~\cite{roussel2011understanding}. Until this point, cement is considered to remain in a flowable state, that can be remixed  and placed. }

Properly characterizing \ar{fresh} concrete thixotropy\ar{, focusing on both the evolution of its yield stress and viscosity with time,} is critical for optimizing its placement, predicting formwork removal time~\cite{tchamba2008lateral}, and overall ensuring quality control in industrial processes \ar{such as additive manufacturing~\cite{roussel2018rheological}}. 
Among several methods to probe rheological properties (from field measurements such as funnel flow and cone spread~\cite{EN12350-8_2019,EN12350-9:2010} to well-controlled lab rheometric tests~\cite{tattersall1983rheology,roussel2011understanding}), the study of sedimenting spheres in cementitious pastes may offer insights into their evolving microstructure and yield stress, and is inspired by work in other thixotropic fluids~\cite{beris1985creeping,gueslin2006flow,ferroir2004motion}. 
The local flow induced by a sedimenting ball induces a local fluidization of the paste, that may or may not be enough to allow its further movement~\cite{biswas2021quantifying,ferroir2004motion}. \ar{This method is particularly interesting, in our viewpoint, for two main reasons. First, fresh cement paste and fresh concrete may localize shear during rheological flow tests, a phenomenon attributed to shear-induced particle migration of cement particles or aggregates, and that may introduce artifacts during rheological tests and compromise the reliability of the properties identified from testing~\cite{haist2020interlaboratory,dhar2024discrepancies,baumert2020minimization,izadi2021squeeze,kabashi2023shear}. In contrast to conventional rheometry, where the same material confined within the measuring gap is repeatedly sheared, a settling sphere continuously traverses previously undisturbed material, mitigating the creation of layers less rich in particles that would localize shear. 
Second, another mechanical loading may be superposed to the local shear imposed by the settling sphere. Among these, vibration and orthogonal shear have been shown to induce particle sedimentation in yield stress~\cite{ovarlez2012shear}, thixotropic~\cite{ferroir2004motion}, and viscoelastic~\cite{gheissary1996unexpected,padhy2013effect} fluids, as well as in dense granular suspensions~\cite{van2019ultrasonic}.
Vibration is used in almost every construction site today to place conventional concrete, allowing it to flow~\cite{tattersall1983rheology}. This is especially critical around steel reinforcement bars, at the risk of seriously compromising elements durability if the concrete does not encapsulate the reinforcement~\cite{thrane2012modelling}.  Such vibrators have been empirically designed and are very effective.  Despite some recent studies~\cite{li2026experimental,zhao2024review}, their effectiveness relies on experience with conventional concrete, an inappropriate application time may lead to defects (trapped air or on the opposite sedimentation of aggregates),
and the vibration intensity needed to place more sustainable concretes~\cite{van2018concrete,juenger2019supplementary,scrivener2018calcined}
is unclear. The current work establishes a baseline for more advanced experiments where a second load will be superposed.}

The settling of particles within non-Newtonian fluids has been extensively studied, due to its relevance in industrial and natural processes~\cite{chhabra2023bubbles}. In Newtonian fluids \ar{and at low particle Reynolds number}, Stokes' law predicts terminal velocity based on fluid viscosity and solid-fluid density difference~\cite{stokes1851effect}. However, yield-stress fluids deviate significantly from this law: particles either settle continuously when the gravitational stress exceeds the yield threshold, or arrest entirely once the fluid structural strength dominates~\cite{ferroir2004motion}. The critical yield stress for particle arrest is derived from force balances, incorporating particle size, density contrast, and fluid rheology~\cite{beris1985creeping,tabuteau2007drag,van2019ultrasonic}. Experimental studies using spheres, cylinders, or disks have revealed complex velocity profiles, transient regimes, and wall effects~\cite{mohammad2023experimental,mrokowska2019viscoelastic,koochi2023interaction,yazdi2023sedimentation,gueslin2009sphere}. Advanced imaging techniques, such as high-speed optical imaging~\cite{biswas2021quantifying}, ultrasonic tracking~\cite{van2019ultrasonic}, X-ray computed tomography~\cite{strybny2024mechanisms} and magnetic resonance imaging (MRI)~\cite{ovarlez2012shear,kleinhans2008magnetic}, have enabled real-time tracking of particle (or bubble) trajectories, providing spatial and temporal resolution.
The settling of spheres in aging yield stress fluids~\cite{gueslin2006flow,tabuteau2007drag,gueslin2009sphere} and thixotropic fluids~\cite{ferroir2004motion,ferrari2021fully,biswas2021quantifying} has been the focus of several studies where paste structuration and paste fluidization are considered in models that can capture, at least partially, the sedimentation of particles in such complex fluids.

In this work, we explore yet another complex fluid, a cement paste, whose rheological behavior displays thixotropy, with rapid structuration due to the flocculation of cement particles~\cite{roussel2012origins,dhar2024discrepancies}.
To do so, the settling of spherical steel balls in a quiescent cement paste
is investigated through X-ray computed tomography (X-ray CT), and the ball velocity computed using image analysis.
The test campaign focuses on
the influence of cement paste age at rest. 
In parallel, the viscoelastic characterization of the paste is performed through small amplitude oscillatory shear, quantifying the evolution of the paste elastic modulus and static yield stress with paste age \ar{at rest}.
Finally, thixotropic models of increasing sophistication are evaluated and their ability to properly capture the paste behavior discussed. 

\ar{This work aims at developing a method to study the rheological properties of cement paste that minimizes artifacts linked to localized shear, and at quantifying the material properties that dictate ball falling velocity and ball stoppage.
This understanding is critical to support the design of sustainable concretes where aggregates do not settle, but where vibration
induces flow. }

\section{Materials and Methods}

\subsection{Sample preparation}
\label{section:Sample_preparation}

Cement paste samples were prepared using Der Contragress cement (42.5 R, $C_3A$-free) in water, with a water-to-cement ratio (w/c) of $0.50$ \ar{(corresponding to an initial cement particle solid volume fraction of 39\%)}, with $0.075\% $ by weight of cement of polycarboxylether superplasticizer (Master Glenium ACE 430 from Master Builders), denoted SP, and $0.075\%$ by weight of cement of retarder. 
The density of the cement paste was calculated from recipe using a density of $3150~\rm kg/m^3$ for the cement particles \cite{thomas2009materials} and of $1000~\rm kg/m^3$ for the water, neglecting the additives, and amounts to $1835~\rm kg/m^3$. 
\\
For the settling ball experiments, batches of two liters of paste were mixed for each experiment. 
SP and retarder were pre-mixed in water and gradually added to the cement during mixing. Mixing was performed with an  IKA\textsuperscript{\textregistered} STARVISC 200-2.5 control overhead U-shaped stirrer at 800 rpm in two steps. First, the paste was mixed for 6 minutes, slowly adding water. Next, the paste was scraped from the sides of the container, and the mixture allowed to rest for 10 minutes. After 10 minutes, it was mixed again for 2 minutes at the same rpm. The paste was then immediately poured into the pipe designed for the settling experiments, and a timer started, with time elapsed described in the rest of the article as \ar{"cement paste age at rest"}.
For rheology tests, cement paste samples were prepared in smaller amounts (80 ml) using the same IKA mixer. The 2-minute-remix was done within the cup of the vane geometry (at 800 rpm), after the same 10 minute rest time as above. 

\ar{Variability in the rheological response of the paste may stem from the amount of paste mixed, especially for the properties measured under small amplitude oscillatory shear, linked to the very small strain applied and to the potential development of heterogeneities in the microstructure \cite{ramachandran2010particle,ovarlez2011flows}.
The choice of geometry also introduces an uncertainty of order of magnitude 10\% on the $G'$ values, partially coming from the estimation of strain in the vane geometry \cite{patarin2014vane,servais2002oscillating,vadillo2025vane}.
Despite its disadvantages, the choice of vane geometry follows our previous flow study on cement paste that shows that the parallel plate geometry introduces an even higher uncertainty (in some case, two orders of magnitude) \cite{dhar2024discrepancies}.
Overall, the time evolution of the elastic modulus carries a significant uncertainty (on the order of 10\%) that is difficult to eliminate. 
}\\
\arr{Throughout this work, time zero is the end of the last high-shear step, and the time elapsed since then is referred to as the paste's "age at rest".}

\subsection{Rheological measurements} 

The rheological measurements were performed using an Anton Paar MCR 302 rheometer with a vane (4 blades, diameter 21.81 mm) in a sanded cup geometry (diameter 28.92mm), described in \cite{dhar2024discrepancies}. The temperature was set to 25\,$^{\circ}$C to match the temperature inside the CT scanner. 

\begin{figure}[!h]
\centering
\includegraphics[width=25pc]{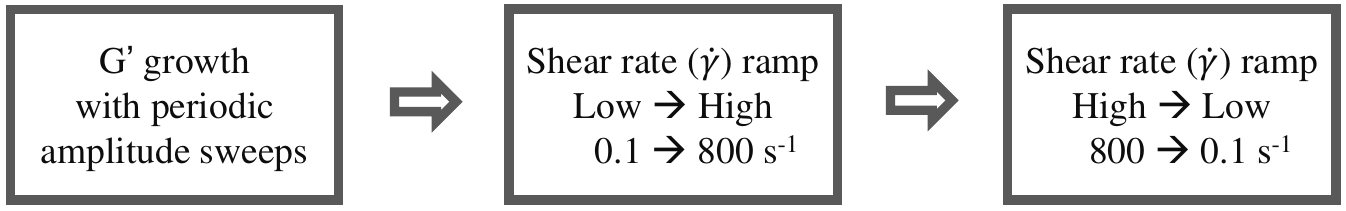}
\captionsetup{width=30pc}
\caption{Rheology protocol used to study the cement paste.} 
\label{fig:rheology_protocol}
\end{figure}

The rheological protocol, shown in Fig. \ref{fig:rheology_protocol}, consisted of three steps, and was applied right after the remix phase of the paste, described in \ref{section:Sample_preparation}. 

\begin{figure}[!h]
\centering
\includegraphics[width=17pc]{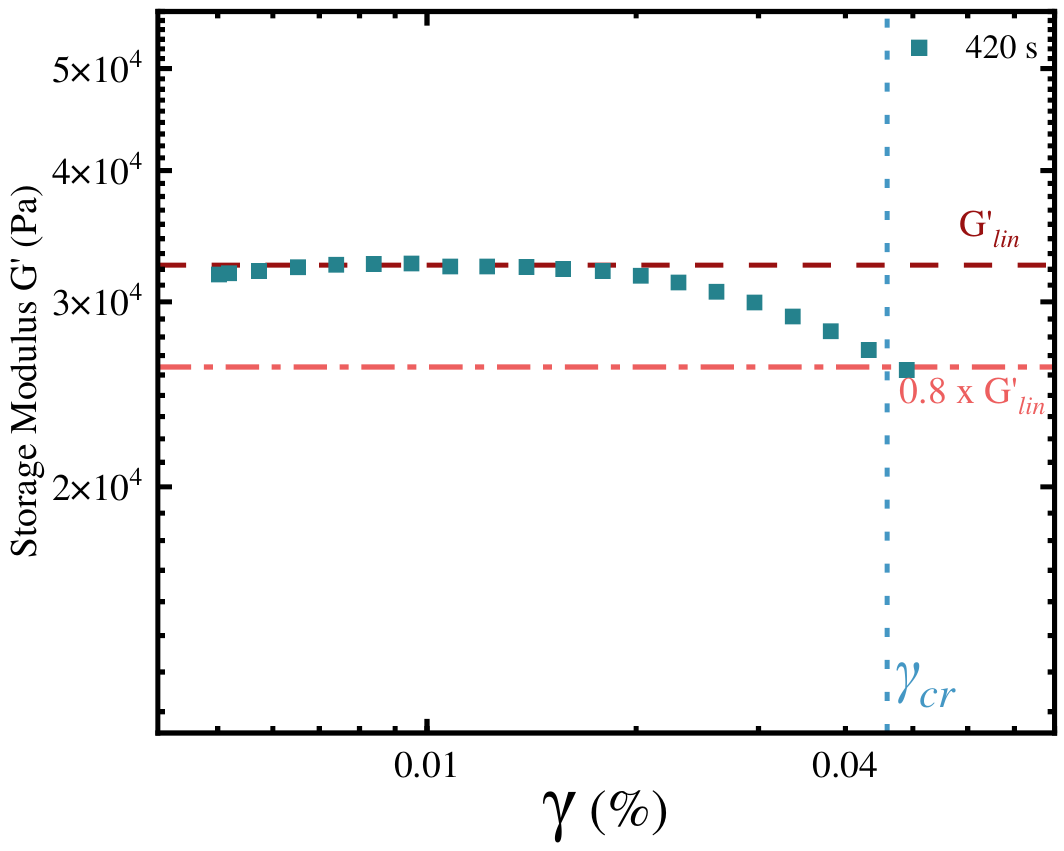}
\captionsetup{width=30pc}
\caption{Strain amplitude sweep conducted at paste age at rest\,=420\,s. It illustrates how $G'_{lin}$ and $\gamma_{cr}$ are determined.} 
\label{fig:SYS_build}
\end{figure}

During the first step,
a small amplitude oscillatory shear of $\gamma = 5 . 10^{-5}$ at a frequency of 1 Hz was applied for a duration of 600 s, and used to follow the storage modulus $G'$ as a function of time. Within this period and at intervals of 60 s, amplitude sweeps up to a strain amplitude of $\gamma = 5. 10^{-4}$ were imposed to the paste. This step enables the estimation of an elastic modulus $G'_{lin}$, taken as the average of the storage moduli in the linear elastic region, and the determination of a critical strain $\gamma_{cr}$, value of the strain amplitude at which the value of $G'$ decreases to 80$\%$ of $G'_{lin}$ (see Fig. \ref{fig:SYS_build}). 
This step is conducted while minimally altering the microstructure of the paste, a hypothesis confirmed when tests without the amplitude sweeps were performed, and showed a similar $G'$ evolution with time (see Fig. \ref{fig:G_growth}).
From $G'_{lin}$ and $\gamma_{cr}$, a static yield stress $\tau_s$= $\gamma_{cr}$\,.\,$G'_{lin}$ was determined as a function of paste age \ar{at rest}. \ar{This definition of static yield stress has been used in cement \cite{nachbaur2001dynamic,roussel2012origins,yuan2017measurement} and is inspired by works on colloidal silica gels \cite{walls2003yield}.}

The second step consisted of a shear rate ramp from low to high, and is used to fluidize the sample. \ar{The applied shear rate was increased from 0.1 $s^{-1}$ to 800 $s^{-1}$ using forty logarithmically spaced shear rate increments generated by the rheometer software. The first step at 0.1 $s^{-1}$ lasted 5 s and the last step at 800 $s^{-1}$ lasted 1 s. The total duration of the step was 100 s. }
The shear rate equivalent values are proposed by Anton Paar \ar{for the vane geometry} and given equal to the rotational velocity in rpm. 

The third step, the ramp from high to low shear rate, is used to measure the apparent dynamic viscosity of the paste as a function of shear rate. \ar{The reverse protocol is applied, with the same logarithmic ramp conditions.}\\
The plateau value between 400 and 800 $s^{-1}$ is chosen to describe the fully fluidized paste shear viscosity, and a Herschel-Bulkley model is fitted on the data to identify the paste dynamic yield stress.

\ar{To assess the reversibility of the paste rheological behavior, or in other terms, to isolate the influence of the irreversible hydration reaction from the reversible paste structuration, the paste $G'$ growth and dynamic yield stress were evaluated at different ages of paste since first contact with water. Results are briefly described in the result section and detailed in supplementary information (SI). This point is also discussed in literature  and it is generally accepted that cement reversible behavior dominates for the first hour since contact with water \cite{roussel2012origins}.}

\subsection{Falling ball setup}

\begin{figure}[!h]
\centering
\includegraphics[width=12pc]{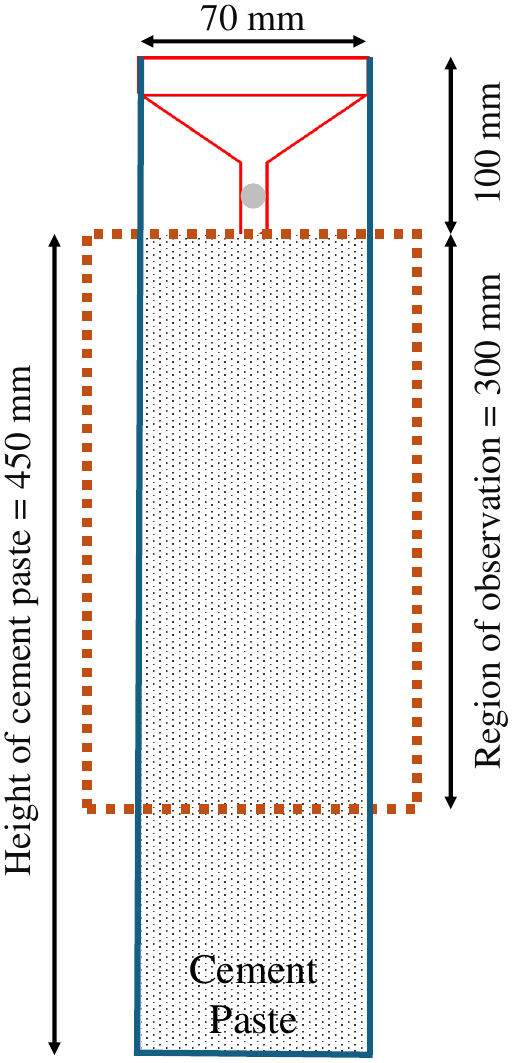}
\captionsetup{width=30pc}
\caption{Schematic of experimental setup inside an X-ray chamber. The tube has a height of 550\,mm and a diameter of 70\,mm, with a 100\,mm funnel and a 300\,mm observation region. The paste is filled up to a height of 450\,mm.} 
\label{fig:experimental_setup}
\end{figure}

\ar{The custom-made setup consisted of a PVC tube of 70 mm diameter and height of 550 mm, fixed on a wooden board to reach a verticality of $\pm$ 0.5\,$^{\circ}$. A schematic is shown in Fig. \ref{fig:experimental_setup}.} 
\ar{PVC} was chosen to mitigate X-ray absorption and the pipe diameter to reach a tube-to-ball diameter ratio of at least 10.

\ar{The ball drop mechanism consisted of an electromagnetic system which holds the steel ball when voltage was applied, and let fall upon breaking the circuit, enabling the ball drop when the sample is inside the X-Ray chamber. A funnel system was added below the electromagnetic system, 
ensuring that the ball enters the paste near the exact center of the pipe. } 
The distance between the ball attached to the magnet and the surface of the cement slurry was about 70 mm. The ball therefore entered the paste with a non-negligible velocity, which was useful for overcoming the paste "skin" - a layer of drying cement paste that may form, especially when the cement paste was left at rest for longer times -. \ar{This initial non-zero velocity ensured} that the ball \ar{could} penetrate the bulk of the paste. \ar{The velocity of the ball decreases to reach a steady state value at a maximum of 60\,mm below the paste surface, far from the depth at which average steady-state velocities were determined (between 100 and 225\,mm below paste surface). }

The CT scanner was UniTOM XL from TESCAN.
 Optimizing the CT scan resolution and the distance to the sensor led to a window of observation of 300 mm height, so that the ball falling could be observed from the cement-air top interface to about 300 mm penetration depth.
 Three different steel ball sizes of radii $R =$\,2, 2.5 and 3 mm were used. 
 
 The steel ball (density $7807~\rm kg/m^3$) was released at different ages of the paste, between 120\,s (the minimum time to pour the cement paste in the tube and place it in the X-Ray chamber) and 480\,s, time at which the biggest ball stopped settling during the experiment, at increments of 60\,s. 
 
 For each ball size and age of paste, a new freshly mixed paste was used (no successive ball falling in a single sample was explored here), and the experiment was repeated between 2 and 5 times, depending on the test reproducibility. The ratio of the tube-to-ball diameter was $\approx$ 12 for the biggest ball, allowing us to neglect wall effects \cite{ladenburg1907einfluss,fidleris1961experimental}.
Bottom wall effects could also be neglected, as observations were made at least 150 mm above the cell bottom. 

Results are shown for balls whose velocity could be resolved at the frequency of acquisition (25 images per second), e.g., the ball of radius 3\,mm  was sedimenting too fast in the paste with 120 s aging time.
When ball stoppage was observed, it was typically preceded by a decrease in its velocity. If the ball was observed to stop within the  observation window, its velocity is counted as zero.
Error bars were calculated as data standard deviation, considering all experiments. 

\ar{Experiments were performed on cement pastes of increasing resting age, with each test labeled according to the paste age at rest at the moment the ball was dropped. For example, the test labeled 240\,s for the R\,=\,2\,mm ball corresponds to a paste with a resting age of 240\,s at the start of the test; as the test lasted $\approx$ 20\,s, the resting age reached $\approx$ 260\,s by the end of the test.}

\subsection{Ball tracking and velocity calculation} 

The set of images recorded during the experiments was optimized for brightness and contrast
using ImageJ  \cite{schneider2012nih}. The ball was then automatically localized using video spot tracker software from CISMM \ar{and its position extracted frame by frame ($\Delta t=0.04$\,s, 25\,fps)}. 
The surface of the cement paste is taken as zero value of the vertical axis. \ar{Finally, the
instantaneous velocity was computed with a central finite-difference scheme,
$v_i(t_k)=[y(t_{k+1})-y(t_{k-1})]/(2\Delta t)$. }

\section{Rheological characterization results}

\subsection{Growth of storage modulus $G'$ and complex viscosity $\eta^*$ as a function of paste age}

\begin{figure}[!h]
\centering
\includegraphics[width=20pc]{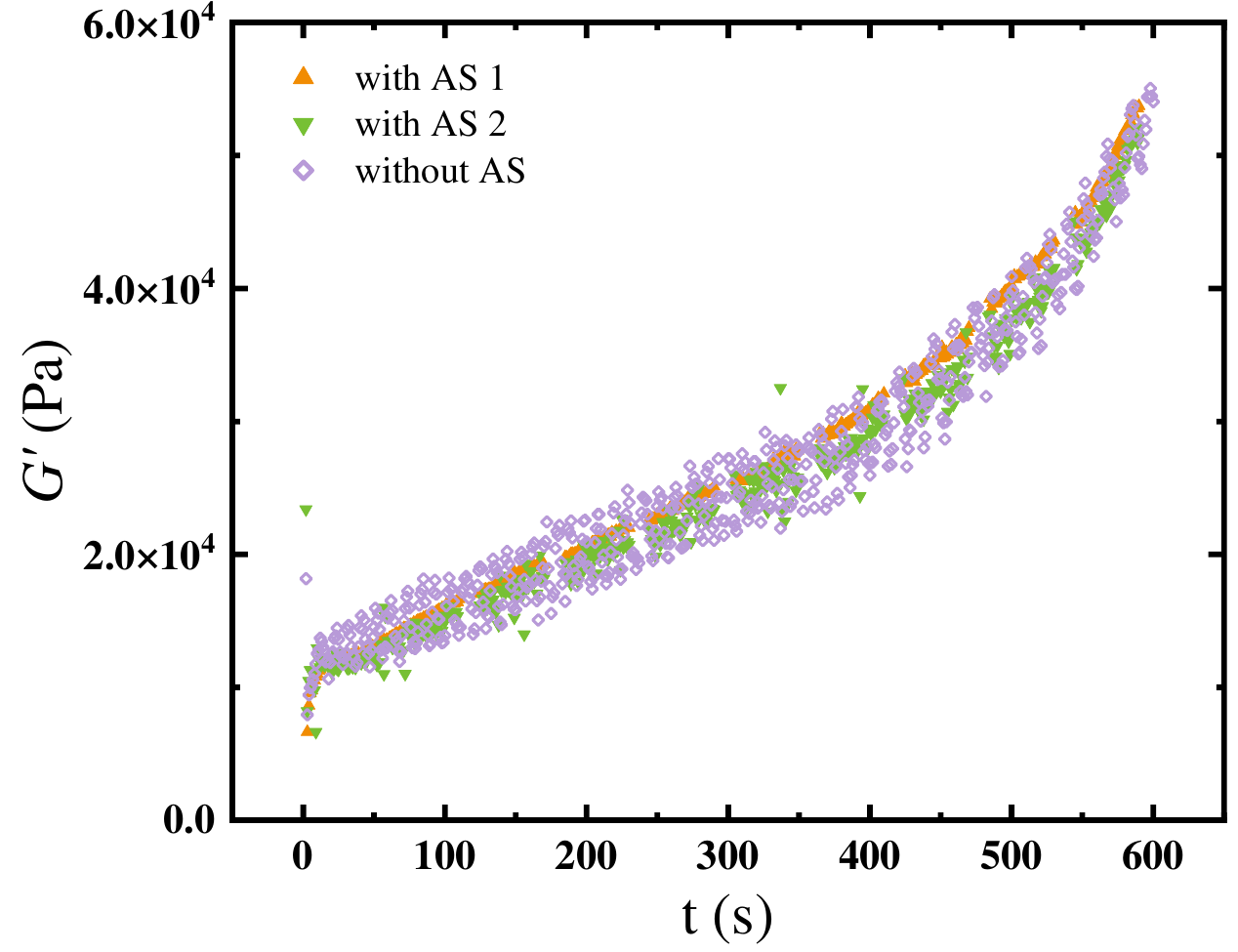}
\captionsetup{width=30pc}
\caption{Growth of storage modulus $G'$ with time (or paste age \ar{at rest}). The trials with amplitude sweep (AS) are shown with orange upward triangles and green downward triangles. The trial without any amplitude sweep is shown by purple diamonds.} 
\label{fig:G_growth}
\end{figure}

Fig. \ref{fig:G_growth} shows the growth in storage modulus $G'$ as a function of time (or paste age \ar{at rest}), obtained from the protocol with regularly imposed amplitude sweeps (2 independently mixed samples), and from the protocol without. \ar{Data are superposed,} confirming that the amplitude sweep minimally disrupts the paste, and its structuration with time at rest.\\
\ar{To evaluate further the reversibility of the paste structuring behavior at rest, a longer protocol consisting of $G'$ growth at rest (600 s), high shear (120 s), and $G'$ growth at rest (600 s) was implemented. Results displayed in SI (Fig. 13) show that both $G'$ growth steps are nearly superposed. In other words, the hydration irreversible time dependent effects are not dominant in the time frame explored in this work. }

\begin{figure}[!h]
\centering
\includegraphics[width=20pc]{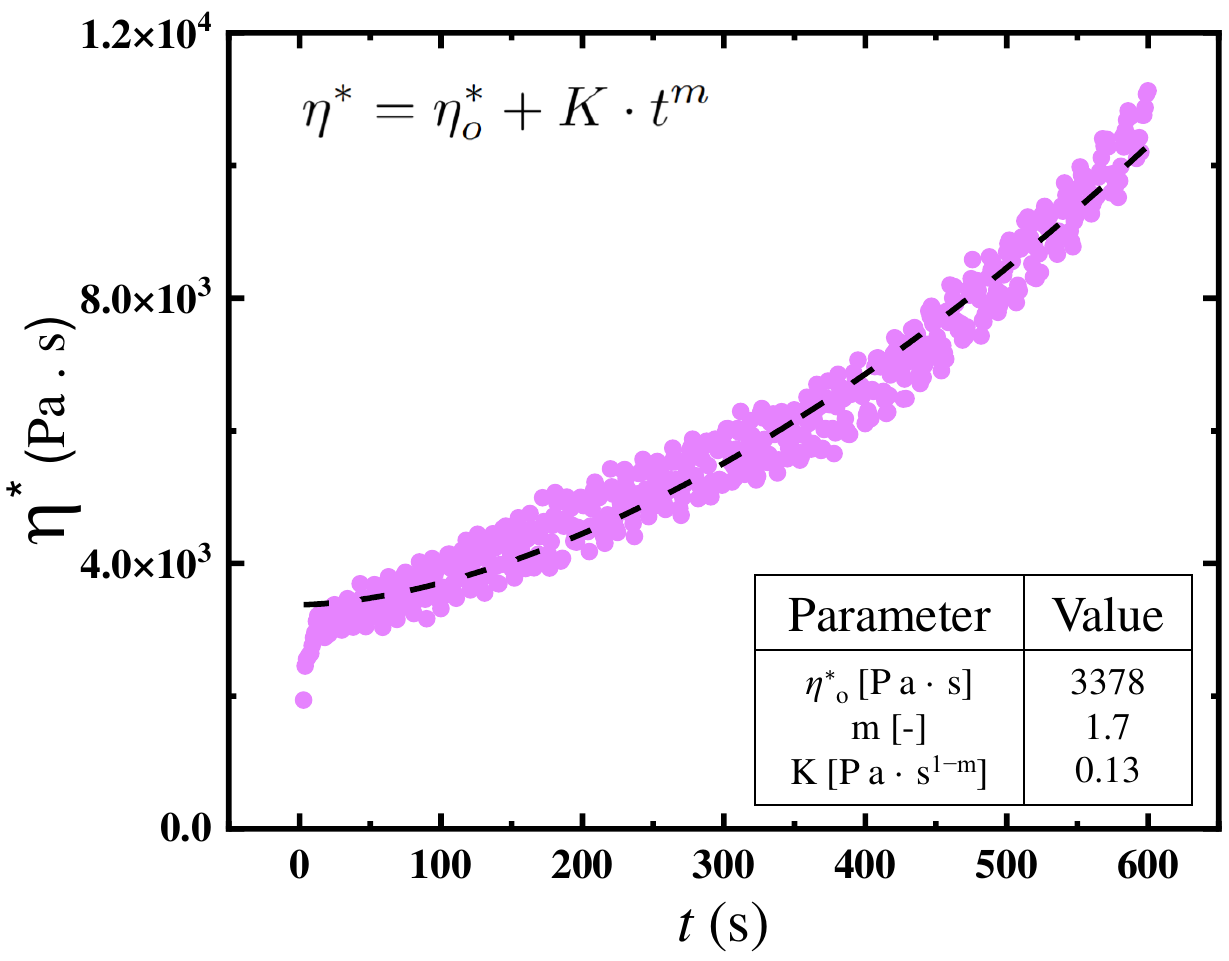}
\captionsetup{width=30pc}
\caption{Growth of complex viscosity $\eta^*$ as a function of time $t$ (or paste age \ar{at rest}) represented by purple dots. The dashed line shows the fit of the equation $\eta^* = \eta_o^* + K\cdot t^m$.} 
\label{fig:complex_viscosity}
\end{figure}

From the measurements of storage and loss moduli, resp. $G'$ and $G''$, the complex viscosity $\eta^*= \frac{\sqrt{G'^2 + G''^2}}{\omega}$, with $\omega$ the angular frequency \cite{hackley2001guide}, was evaluated, and is plotted in Fig. \ref{fig:complex_viscosity} as a function of time. 

The data can be fitted by a simple equation, that captures the paste structuration, with the form:

\begin{equation}
 \eta^* = \eta^*_o + K\cdot t^m   \label{eqn:complex_visc}
\end{equation}
\hspace{15em}

 where $\eta^*_o$ is the complex viscosity of the paste at time t=0, and $K$ and $m$ are material parameters.
\\
The fit of Eqn. \ref{eqn:complex_visc} on experimental data returned the parameters shown in the inset of Fig. \ref{fig:complex_viscosity}.
\arr{As mentioned in section \ref{section:Sample_preparation}, the storage modulus data carries a significant uncertainty that is propagated to the complex viscosity and identified parameters of the equation.}

\subsection{Amplitude sweep and evaluation of the growth of static yield stress $\tau_s$ with time}

\begin{figure}[!h]
\centering
\includegraphics[width=30pc]{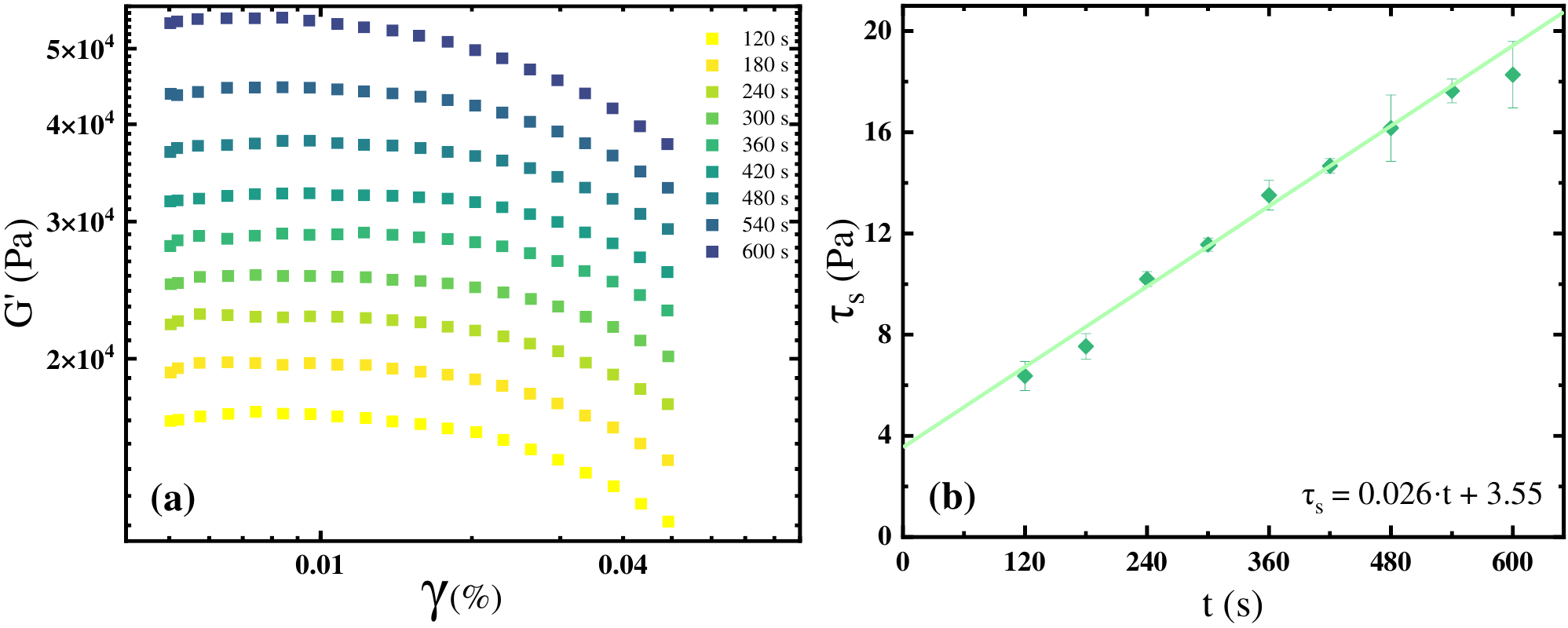}
\captionsetup{width=30pc}
\caption{(a) Amplitude sweep at different \ar{paste ages at rest} showing the drop of storage modulus $G'$ as a function of increasing strain amplitude $\gamma$. The gradient of color from light yellow to dark purple depicts increasing paste age at rest from 120 to 600 s. (b) Growth of static yield stress $\tau_s$ as a function of \ar{cement paste age at rest} $t$ (green diamonds). The error bars represent the range (maximum-minimum) for two independent trials. The fitted green line shows a linear fit and returns $\tau_s [Pa]= 0.026\cdot t [s] + 3.55$.} 
\label{fig:AS_static_YS}
\end{figure}

Amplitude sweeps, performed every 60\,s \ar{during the $G'$ growth experiments} 
between 120 and 600\,s, are shown in Fig. \ref{fig:AS_static_YS}(a). \ar{The time shown in legend corresponds to paste age at rest}.

\ar{The procedure to estimate the static yield stress $\tau_s$ from $\gamma_{cr}$ and $G'_{lin}$ (described with Fig. \ref{fig:SYS_build}) was applied to all curves, and $\tau_s$ could be determined for increasing values of paste age at rest. As a result,} the evolution of the  \ar{paste} static yield stress  $\tau_{s}$ 
\ar{was plotted as a function of paste age at rest and} is shown in Fig. \ref{fig:AS_static_YS}(b). 

The growth in $\tau_{s}$ \ar{appears linear with time, and} was fitted with \ar{the following} 
equation \cite{roussel2006thixotropy}: 

\begin{equation}
\label{eqn:tau_fit}
\tau_{s}(t) [Pa]= 0.026\,[Pa.s^{-1}]\cdot t \,[s] + 3.55 \,[Pa]
\end{equation}

\subsection{Viscosity evaluation }

\begin{figure}[!h]
\centering
\includegraphics[width=20pc]{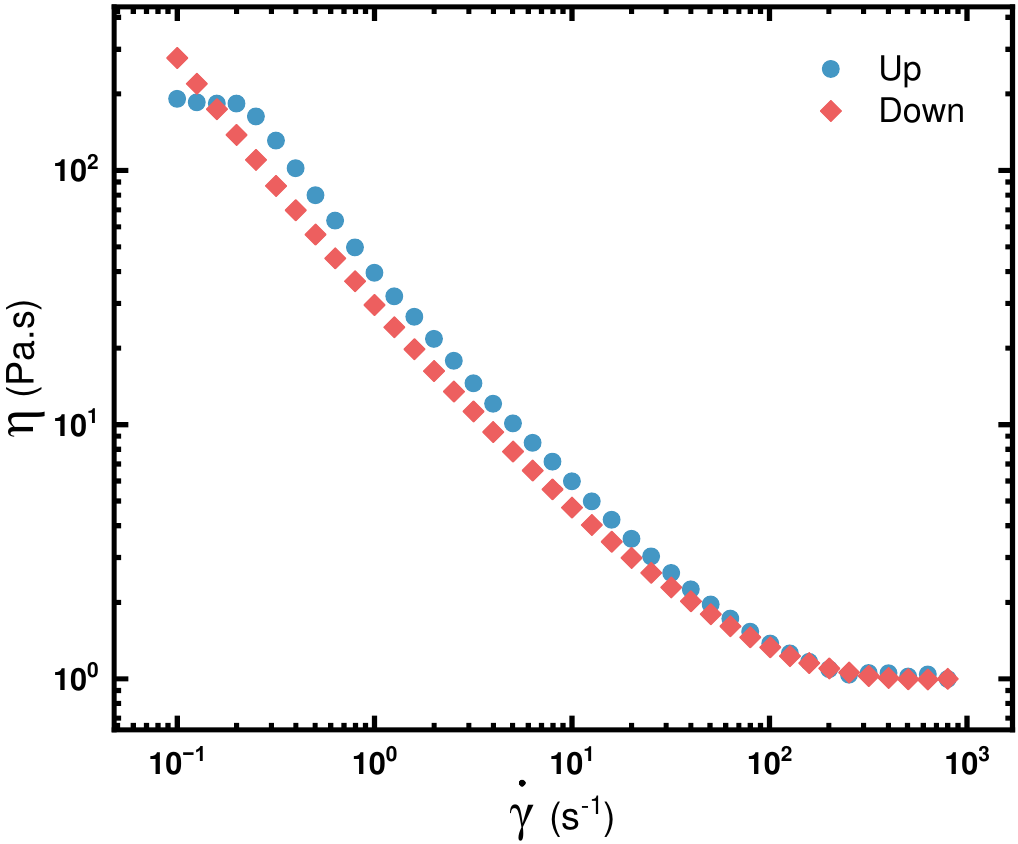}
\captionsetup{width=30pc}
\caption{Cement paste viscosity as a function of shear rate from flow test} 
\label{fig:FlowCurve}
\end{figure}

\ar{The shear ramp results are plotted in Fig. \ref{fig:FlowCurve}. }
The viscosity of the paste is shown to plateau between 400 and 800 $s^{-1}$ to reach $1\, \pm 0.1\,Pa.s$, \ar{a value that is considered representative of the paste in its fully-fluidized form.} 


\subsection{Dynamic yield stress evaluation }

\begin{figure}[!h]
\centering
\includegraphics[width=20pc]{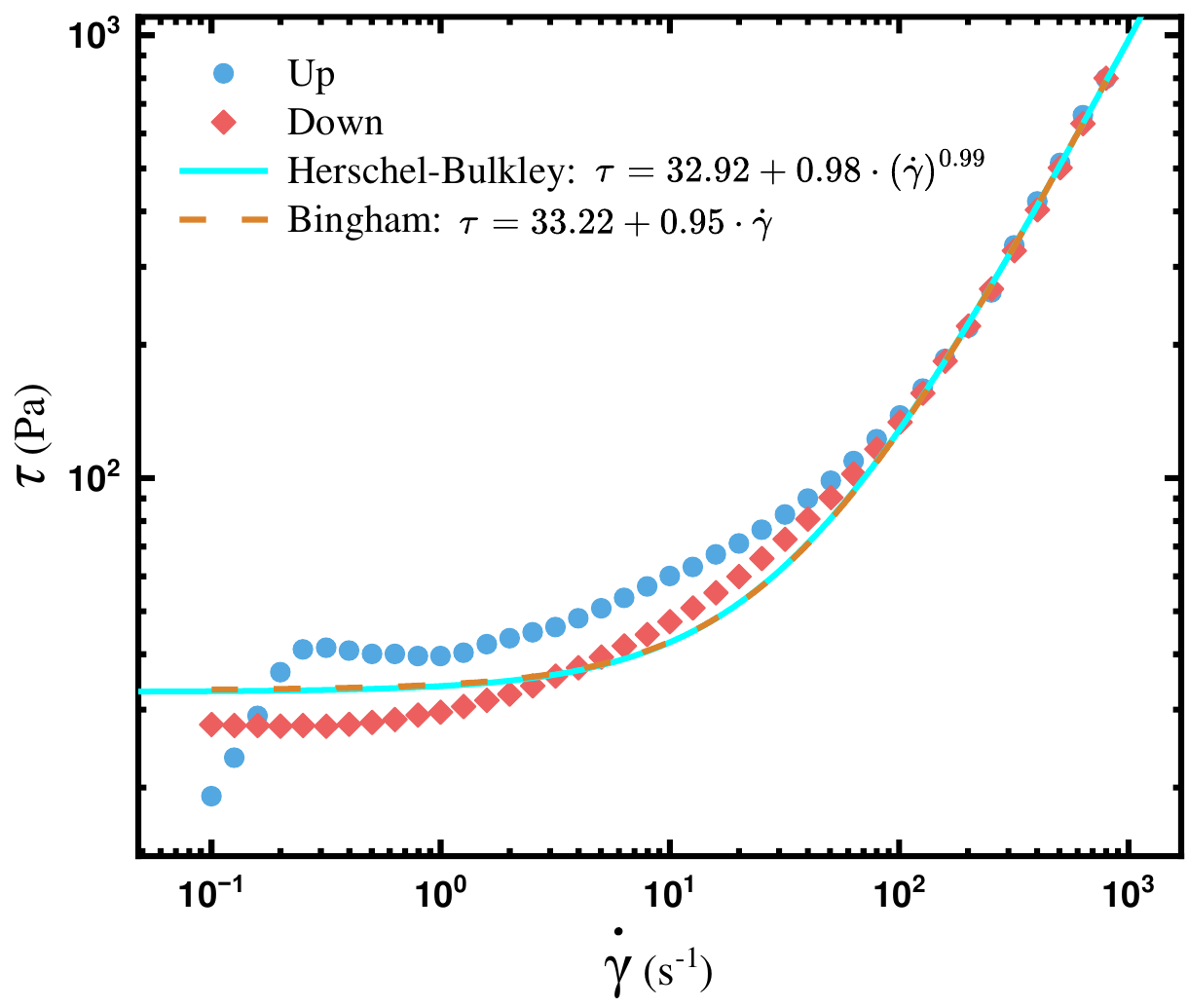}
\captionsetup{width=30pc}
\caption{Cement paste shear stress $\tau$ as a function of shear rate $\dot\gamma$, with Herschel Bulkley (cyan) and Bingham (orange) fits to identify dynamic yield stress $\tau_o$, from ramp down.} 
\label{fig:Dyn_YS}
\end{figure}

\ar{The shear ramp from high to low was used to identify the paste dynamic yield stress using a Herschel-Bulkley model fit, leading to  33 Pa (see Fig. \ref{fig:Dyn_YS}).}\\
\ar{The difference between the static and dynamic yield stress may be linked to the brittle behavior of the cement paste. For example, Carbopol gel exhibits a $\gamma_{cr}$ (end of the linear visco-elastic domain) of $\approx$ $10^{-2}$ \cite{firouznia2018interaction}, whereas $\gamma_{cr}$  measured on the cement paste is $\approx$ $5.10^{-4}$.}\\
\ar{Here again, we estimated the reversibility of the paste behavior by performing a longer protocol consisting of rest (duration 120 s), shear ramps low-to-high and high-to-low (duration 200 s), shear at 800 $s^{-1}$ (duration 120 s), rest (duration 480 s), shear ramps low-to-high and high-to-low (duration 200 s). The dynamic yield stress identified from both high-to-low ramps (separated by 800\,s duration) differed by less than 15~\% (see SI, Fig. 14), confirming that the hydration irreversible effects are not dominant in the time frame explored in this work}.

\section{Ball settling results: Ball velocity and ball stoppage in the cement paste as a function of its age at rest}

\begin{figure}[!h]
\centering
\includegraphics[width=28pc]{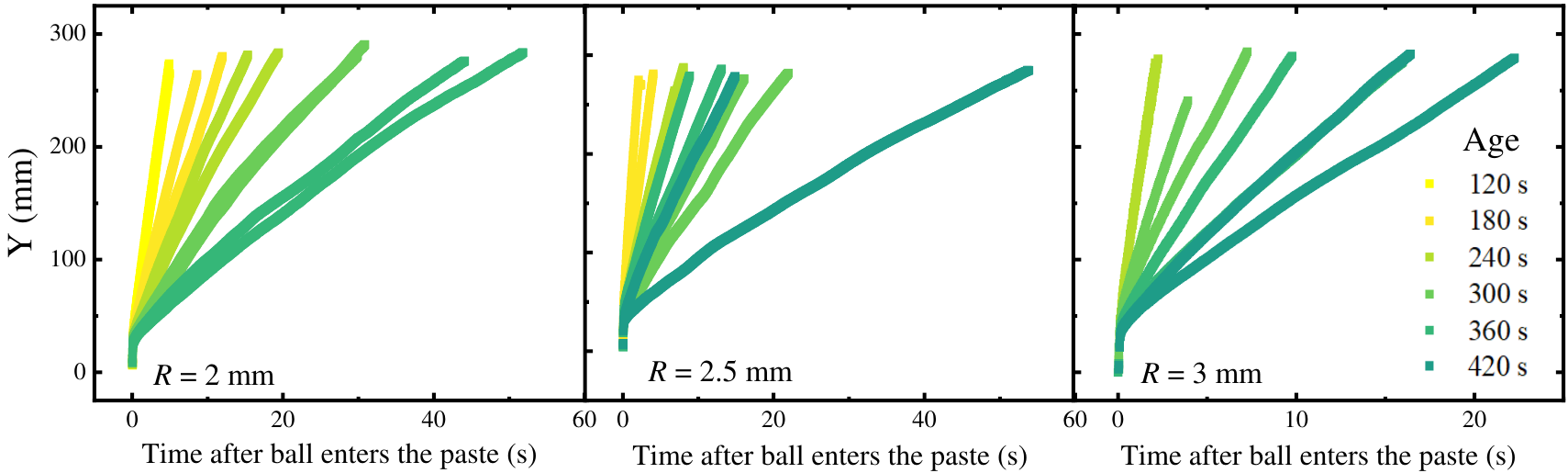}
\captionsetup{width=30pc}
\caption{Measured vertical position of falling ball as a function of time, for various cement paste ages \ar{at rest at the start of settling ball test} and ball radii (2\,mm -left graph-, 2.5\,mm -middle graph- and 3\,mm -right graph-), for 2 independent trials in each condition. Time is here defined as the time elapsed since the ball is first recorded to touch the cement paste. The gradient of color from light yellow to dark green represents increasing \ar{cement} paste age \ar{at rest at the beginning of the test}, from 120 to 420\,s. The time at which the ball stopped within the region of observation is not shown here. } 
\label{fig:position_time}
\end{figure}

The results of ball localization within the 300 mm observation window are shown in 
Fig. \ref{fig:position_time} for the \ar{three} ball sizes R\,=\,2, 2.5 and 3 mm, at cement paste age at rest between 120 and 480\,s, i.e., in all conditions where the ball was moving through the window, and where enough data could be acquired. Test repeats are plotted, showing an acceptable reproducibility of the \ar{data. \\
The tests where ball stoppage was observed within the window of observation are not displayed. In particular, the ball R=2\,mm stopped during test in the paste that was left at rest for 420\,s in all three trials. The ball R\,=\,2.5\,mm stopped in one of three trials (data shown in SI Fig. 16). In some cases, the ball settled too fast for reliable data to be acquired, and results are not displayed. This is the case for the ball R\,=\,2.5\,mm in the paste left at rest for 120 s, and for the ball R=3\,mm in the paste left at rest for 120 s and 180 s.\\
Results show that the time taken for the ball to settle the 300 mm height of cement paste increased with the paste age at rest. Overall, the ball takes between 2.4 and 54\,s to cross the 300 mm of window observation. }

\begin{figure}[!h]
\centering
\includegraphics[width=30pc]{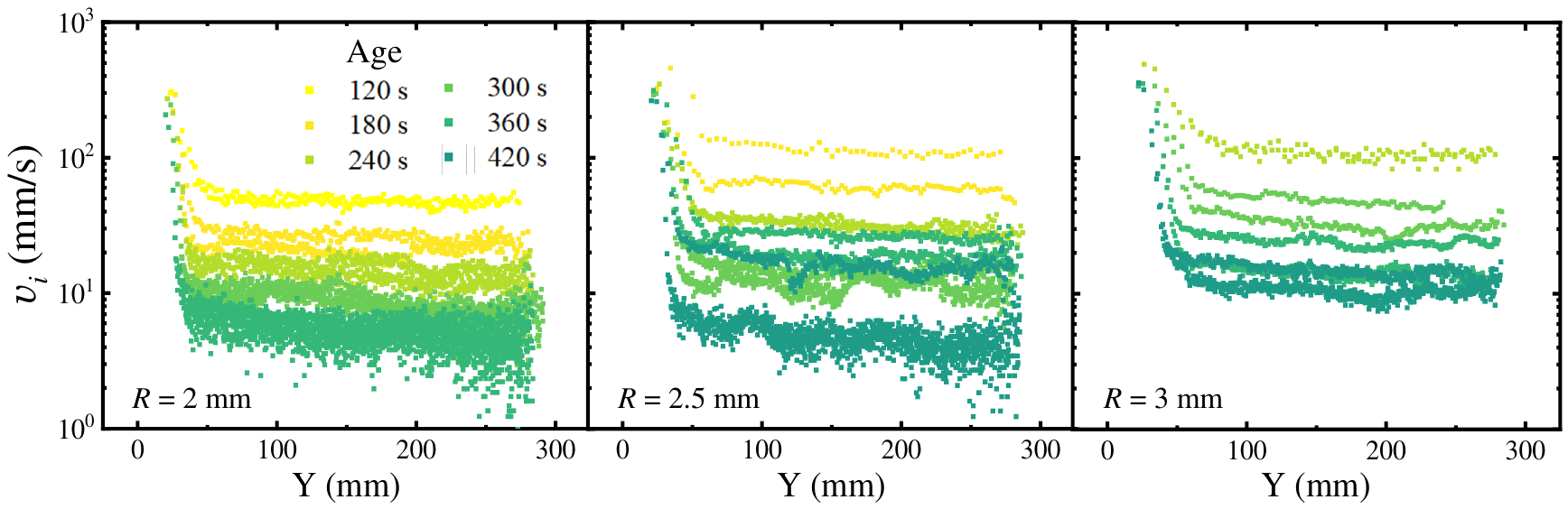}
\captionsetup{width=30pc}
\caption{Measured instantaneous velocities of falling balls $v_i$  as a function of vertical position Y of the ball for \ar{various cement paste ages at rest} and ball radii. The gradient of color from light yellow to dark green represents increasing cement paste \ar{age at rest}, probed between 120 and 420 s. Two test repeats are shown.} 
\label{fig:vel_position}
\end{figure}

The instantaneous velocity $v_i$ of the ball as a function of its position was derived from the above data, and is shown in Fig. \ref{fig:vel_position}.   
After the initial first few seconds, the ball attains a steady-state velocity, where the velocity remains almost constant during the window of observation, \ar{except at higher paste age at rest for the balls of radius 2 and 2.5 mm, where the instantaneous velocity gets noisy and a slight slope is visible on the log scale axis. }

\begin{figure}[!h]
\centering
\includegraphics[width=24pc]{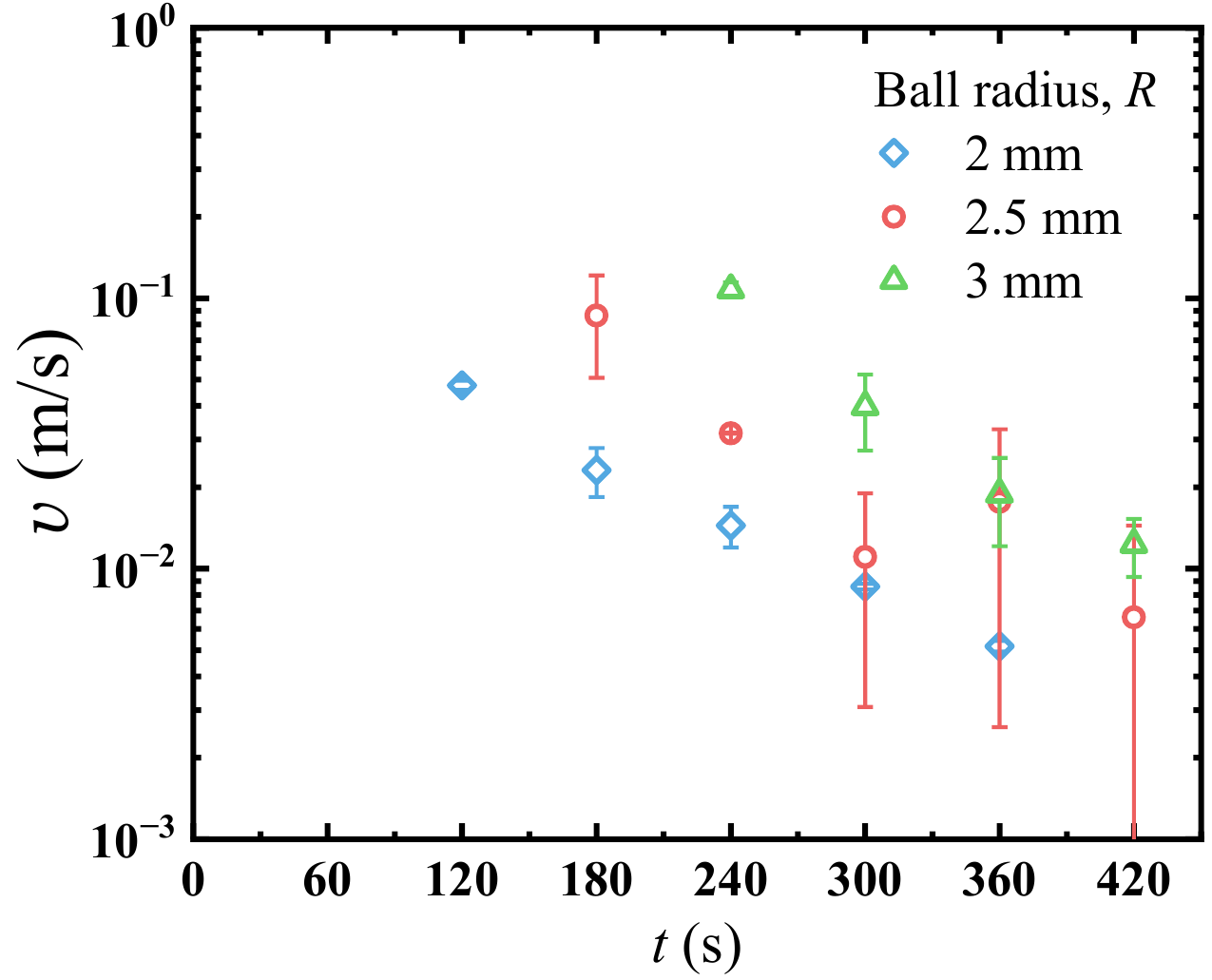}
\captionsetup{width=30pc}
\caption{Experimental steady state velocity $v$ of the falling balls as a function of the paste age \ar{at rest} $t$, in a log-lin scale. Three ball sizes are used with radii 2\,mm (blue diamonds), 2.5\,mm (red dots) and 3\,mm (green triangles). The error bars represent the data standard deviation.} 
\label{fig:steady_velocity}
\end{figure}

The \ar{near} steady-state velocity $v$, calculated by averaging the instantaneous velocity values $v_i$ at ball positions between 100 and 225 mm, \ar{i}s plotted as a function of paste age \ar{at rest} in Fig. \ref{fig:steady_velocity}. The data points corresponding to paste age at which the ball was observed to stop were \ar{not} added \ar{on this graph}.
The data shows that $v$ decreases logarithmically with paste age, until the ball stops. Similar trends were observed on aging Laponite suspensions \cite{gueslin2006flow,gueslin2009sphere}.
\ar{This trend cannot be directly compared to the measured static yield stress as a function of paste age at rest, observed to evolve linearly with time.}

The particle Reynolds numbers were estimated using the measured ball settling velocities and the equation for the average shear rate around the sedimenting ball $\dot{\gamma}=v/2R$ (with $v$, ball settling velocity and $R$ ball radius) \cite{ovarlez2012shear}, and remain in all cases below 0.4.

\section{Theory: Sedimenting ball in a thixotropic fluid }
\label{section:theory}

\subsection{Theory: Velocity of a sphere falling through a quiescent aging thixotropic fluid}

The terminal velocity $v$ of a sphere falling through a Newtonian fluid is given by Stokes' equation \cite{stokes1851effect}: 

 \begin{equation}
\label{eqn:terminal_velocity_Stokes}
 v = \frac{2}{9} \frac{(\rho_{ball} - \rho_{paste}) g R^{2}}{\eta} 
\end{equation}
\\
With $R$, the radius of the ball, $g$, the acceleration due to gravity, $\rho_{ball}$, the density of the ball, $\rho_{paste}$, the density of the fluid and $\eta$, the viscosity of the fluid. 

The equation is valid for Reynolds number Re$\ll$1, but acceptable for Re$<$1 \cite{Durbin_Medic_2007}.\\

In thixotropic fluids, the viscosity $\eta$ does not remain constant, and can be described as a function of a structure parameter $\lambda$ \cite{coussot2002avalanche}, \ar{that tracks the degree of microstructure of the paste and can vary between 0 (fully destructured) and 1 (fully structured)}:

\begin{equation}
 \eta(\lambda) = \eta_{o}(1 + \lambda^{n})
\end{equation} 
\\
where $\eta_{o}$ is the viscosity of the fluid in its fully fluidized form (or fully de-structured, i.e., when $\lambda=0$), and the exponent $n$, a material parameter.\\

\ar{Following the work of Ferroir and co-authors \cite{ferroir2004motion}, t}he structure parameter $\lambda$ \ar{is} split into a structural growth parameter $\lambda_{g}$ and a breakdown parameter $\lambda_{b}$, independent of each other, giving

\begin{equation}
\label{eqn:lambda_total}
 \lambda = \lambda_{g} + \lambda_{b}
\end{equation}

 The new equation for viscosity can then be introduced back into the Stokes' equation to obtain the terminal velocity of a sphere falling through the thixotropic material \cite{ferroir2004motion}, giving:

\begin{equation}
\label{eqn:terminal_velocity}
 v = 
W[1 + (\lambda_g + \lambda_b)^{n}]^{-1}
\end{equation}\\

with \begin{equation} W = \frac{2}{9} \frac{(\rho_{ball} - \rho_{paste}) g R^{2}}{\eta_{o}}  \end{equation}
\\
In this work, we aim at exploring the aspects of cement paste thixotropy that control the sphere settling velocity, and we therefore consider different expressions of the structural parameter $\lambda$. Using literature, we \ar{evaluate} two different expressions of the structural growth parameter $\lambda_{g}$, and two different expressions of the breakdown parameter $\lambda_{b}$. \ar{The split of  $\lambda$ allows us to examine the aspects of rheology that control the ball settling rather than to find an efficient model for computation or to identify a constitutive equation that minimizes the number of material parameters. We refer to other works on single variable structure kinetics models \citep{coussot2002viscosity,roussel2006thixotropy} or full thixotropic elasto-viscoplastic models \citep{desouzamendes2012critical,larson2019review,varchanis2019modeling}. }

Below, we write the equations to within a constant, i.e., the constants of integration are not displayed as they are in the work that inspired this study \cite{ferroir2004motion}. The expressions are described below:\\

\textbf{Expression 1 for the structure growth parameter $\lambda_{g1}$} 

The structure parameter $\lambda_{g}$ is assumed to grow linearly with time (or paste age) \cite{roussel2006thixotropy}, thus 

\begin{equation}
\label{eqn:lambda_g1}
 \lambda_{g1}(t) = \frac{t}{T}
\end{equation}

with $T$ a characteristic time (unit [s]), property of the material. \\ 

\textbf{Expression 2 for the structure growth parameter $\lambda_{g2}$} \

Here, the growth structure parameter $\lambda_{g2}$ is assumed to change as the complex viscosity $\eta^*$ measured under oscillatory shear \cite{saadat2023rheologist, franck2004understanding, kaya2020rheological}.
The temporal evolution of $\lambda_{g2}$ is then written in the form of the temporal evolution of $\eta^*$ (Fig. \ref{fig:complex_viscosity} and Eqn. \ref{eqn:complex_visc}), 
as 

\begin{equation}
\label{eqn:lambda_g2}
\lambda_{g2} = \frac{K}{\eta_o^*}(\frac{t}{T^{'}})^m
\end{equation}\\

with $T^{'}$ a \ar{ dimensionless constant} (unit [-]), and $\eta_o^*$, $K$ and $m$, the values identified from rheological tests and shown in Fig. \ref{fig:complex_viscosity}.
\\

Similarly, the breakdown structure parameter $\lambda_b$ can be written using two different expressions:\\

\textbf{Expression 1 for the structure breakdown parameter $\lambda_{b1}$} \\
Here, the ball movement through the paste is assumed to affect only a small volume of the paste around it, and its effect is neglected. In other words, the passage of the ball is presumed to have no effect on the paste breakdown: 

\begin{equation}
 \lambda_{b1} = 0
\end{equation}
\\

\textbf{Expression 2 for the structure breakdown parameter $\lambda_{b2}$} 

Following the work of Biswas et al. \cite{biswas2021quantifying}, we use an expression of the structure breakdown parameter that evolves proportionally to the ball diameter and time, and follows

\begin{equation}
\lambda_{b2} = -\varepsilon (2\cdot R) t 
\end{equation}

with $\varepsilon$ a parameter, with unit [$m^{-1}.\,s^{-1}$].

The negative sign indicates the drop in viscosity linked to the breakdown of the paste.\\

Based on the four expressions above, we can write four cases for the expression of the structure parameter  $\lambda$, re-introducing the constant of integration as $\lambda_{o}$. From there, four expressions of the settling ball velocity $v$  are proposed, denoted using Roman numbers:
\\

\textbf{Case I $\lambda_{I}$=$\lambda_{g1}$ + $\lambda_{b1}$:} \\
Here, the growth structure parameter increases with time $t$ as 1/T and the breakdown structure parameter is zero. Hence, the structure parameter becomes:

\begin{equation}
\label{eqn:case_1A}
 \lambda_{I} = \frac{t}{T} + \lambda_{o}
\end{equation}

$\lambda_{o}$ represents the initial structure parameter of the material.
\\

The velocity becomes: 

\begin{equation}
\label{eqn:terminal_velocity_I}
 v_{I} =  W \biggl[1 + {\biggl(\frac{t}{T} + \lambda_{o}}\biggr)^{n}\biggr]^{-1}
\end{equation}
\\

\textbf{Case II $\lambda_{II}$=$\lambda_{g1}$ + $\lambda_{b2}$:} \\

Here, the growth structure parameter increases with time $t$ as 1/T and the breakdown structure parameter increases linearly with time and ball radius $R$. Hence, the structure parameter becomes:

\begin{equation}
\label{eqn:case_1B}
 \lambda_{II} = \frac{t}{T} - \varepsilon (2\cdot R) t +  \lambda_{o}
\end{equation}\\

The velocity becomes: 

\begin{equation}
\label{eqn:terminal_velocity_II}
 v_{II} =  W \biggl[1 + {\biggl(\frac{t}{T} - \varepsilon (2\cdot R) t +  \lambda_{o}}\biggr)^{n}\biggr]^{-1}
\end{equation}
\\

\textbf{Case III $\lambda_{III}$=$\lambda_{g2}$ + $\lambda_{b1}$} \\

Here, the growth structure parameter increases with $t^m$  and the breakdown structure parameter is zero. Hence, the structure parameter becomes:

\begin{equation}
\label{eqn:case_2A}
 \lambda_{III} = \frac{K}{\eta_o^*}(\frac{t}{T^{'}})^m +  \lambda_{o}
\end{equation}

The velocity becomes: 

\begin{equation}
\label{eqn:terminal_velocity_III}
 v_{III} =  W \biggl[1 + {\biggl(\frac{K}{\eta_o^*}(\frac{t}{T^{'}})^m +  \lambda_{o}}\biggr)^{n}\biggr]^{-1}
\end{equation}
\\

\textbf{Case IV $\lambda_{IV}$=$\lambda_{g2}$ + $\lambda_{b2}$} \\

Here, the growth structure parameter increases with $t^m$  and the breakdown structure parameter increases linearly with time and ball radius $R$. Hence, the structure parameter becomes:

\begin{equation}
\label{eqn:case_2B}
 \lambda_{IV} = \frac{K}{\eta_o^*}(\frac{t}{T^{'}})^m - \varepsilon (2\cdot R) t +  \lambda_{o}
\end{equation}

The velocity becomes: 

\begin{equation}
\label{eqn:terminal_velocity_IV}
 v_{IV} =  W \biggl[1 + {\biggl(\frac{K}{\eta_o^*}\bigl(\frac{t}{T^{'}})^m - \varepsilon (2\cdot R) t +  \lambda_{o}}\biggr)^{n}\biggr]^{-1}
\end{equation}
\\

In cases I to IV,   \begin{equation} W = \frac{2}{9} \frac{(\rho_{ball} - \rho_{paste}) g R^{2}}{\eta_{o}}  \end{equation}

\subsection{Theory: Ball stoppage in a yield stress fluid}

Beris and co-authors derived an infinite drag coefficient from the analysis of the yielded and unyielded regions of a sphere settling through a Bingham fluid \cite{beris1985creeping}. It was linked to settling ball stoppage in various yield stress fluids \cite{tabuteau2007drag,ovarlez2012shear}, and led to the inequality (\ref{eqn:beris}), which describes the conditions under which a settling ball ceases to move. 

\begin{equation}
\label{eqn:beris}
    \frac{\tau_y}{(\rho_{ball} - \rho_{paste})\cdot g\cdot 2R}\geq\frac{1}{21}
\end{equation}\\

where $\tau_y$ is the yield stress of the fluid, $\rho_{ball}$ and  $\rho_{paste}$ are the densities of the ball and the paste respectively, and $R$ is the radius of the ball. \\

\section{Model fit on experimental data: Results and discussion}
\label{section:fitting}

\subsection{Model to predict the velocity of balls of various diameters falling through the quiescent aging cement paste}

The experimental values of the settling sphere velocity in cement paste shown in Fig. \ref{fig:steady_velocity} (all three sphere radii) are used to fit the equations developed for case I (linear increase of structure growth parameter with time and no consideration of breakdown), case II (linear increase of structure growth parameter with time, and linear decrease of breakdown parameter with time), case III (non linear increase of structure growth parameter with time and no consideration of breakdown) and case IV (non linear increase of structure growth parameter with time and linear decrease of breakdown parameter with time). \\

For the identification, the following constants were used: $g = $ 9.81 $m/s^2$,  $\rho_{ball} - \rho_{paste}=$ 5972 $kg/m^3$, and $\eta_{o}$ = 1 $Pa\cdot s$ (viscosity of the paste at $\dot\gamma$ $\approx$ $800\,s^{-1}$ (Fig. \ref{fig:FlowCurve})). 
\ar{The material parameters describing the structure growth parameter $\lambda_{g2}$ were identified from the evaluation of complex viscosity with paste age at rest (Fig. \ref{fig:complex_viscosity}), with  $\eta^*_o$= 3378 Pa.s (complex viscosity of the paste at time t=0), $K$=0.13 Pa.s$^{1-m}$ and $m$=1.7 (material parameters). } 
In addition, the initial structure parameter of the material, represented by $\lambda_{o}$, was considered to be near zero as the material was sheared at a high rate just before the start of the experiment, and is considered fully fluidized. For this study, it was arbitrarily set to 0.01 (a small value after heavy mixing, but not zero to prevent convergence issues).

During the fitting procedure, all three remaining material parameters ($n$, the exponent, $T$, a characteristic time, or $T^{'}$, a dimensionless constant, and $\varepsilon$, a proportionality factor) were identified. The results are shown in Tab. \ref{tab:parameter_cases}, with R-square values.

\begin{table}[h!]
\centering
\caption{Parameter values identified on the integrality of the settling experiments}
\begin{tabular}{|c|c|c|c|c|}
\hline
\textbf{Parameter / Case} & \textbf{Case I} & \textbf{Case II} & \textbf{Case III} & \textbf{Case IV} \\
\hline
$n$ [-]& 4.80 & 3.49 & 2.82 & 1.11 \\
$T$ [s] & 220.41 & 131.06 & - & - \\
$T^{'}$ [-]& - & - & 0.56 & 0.20 \\
$\varepsilon$ [$m^{-1}.\,s^{-1}$] & - & 0.59 & - & 4.22 \\
R-square & 0.56 & 0.70 & 0.56 & 0.91\\
\hline
\end{tabular}

\label{tab:parameter_cases}
\end{table}

\begin{figure}[!h]
\centering
\includegraphics[width=30pc]{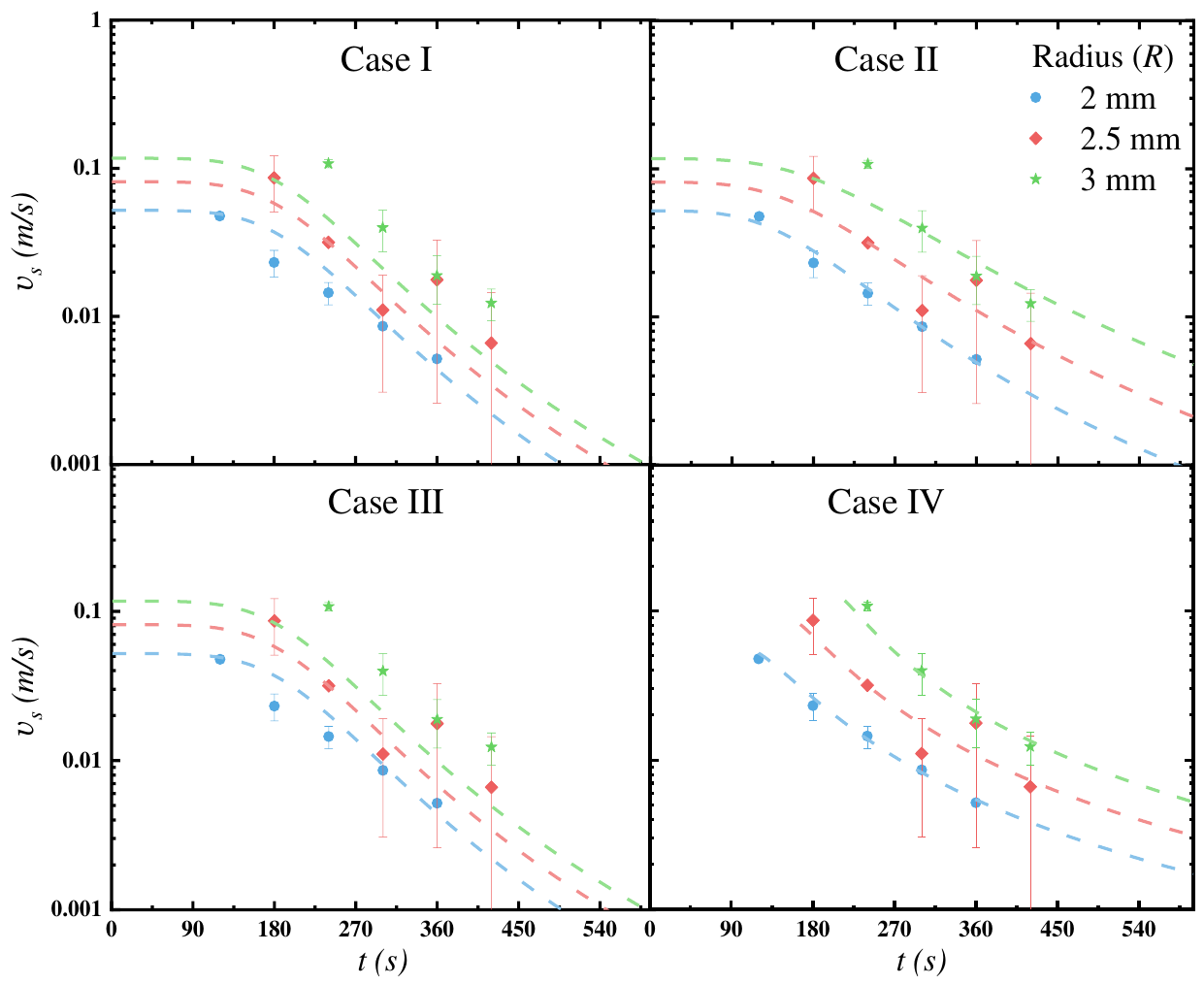}
\captionsetup{width=30pc}
\caption{Experimental (dots) and theoretical (lines) ball settling velocities for the models I to IV, in log-lin scale. Models were fitted with 2 parameters for cases I and III, and 3 parameters for cases II and IV. The error bars represents the data standard deviation.} 
\label{fig:global_fits}
\end{figure}

 Results of the identification are shown on a 2D graph Fig. \ref{fig:global_fits}. The 3D fits of cases I-IV (time, ball radius and velocity) are shown in Fig. 15 of SI.
 The R-square value of 0.91 obtained for case IV shows the importance of considering paste breakdown in relation to shearing of the paste when the ball passes through it.
 \\

\arr{We briefly discuss the identified parameters below. They should be interpreted with caution: the model, while physics-inspired, remains phenomenological, and the parameters come from fitting velocity data rather than from independent measurements of microstructural quantities.}

\begin{itemize}
  \item[-] $n$, the exponent of $\lambda$, governs the sensitivity of the viscosity to the degree of structuration. It is in all cases above 1, i.e., when structuration increases, viscosity increases.
\item[-] $T$, the thixotropic characteristic time, is of order 100\,s and is aligned with literature describing structuration of cement pastes (e.g., \cite{roussel2006thixotropy}). \arr{It is also consistent with the time scale of the measured static yield stress, which doubles from 6 Pa to about 12\,Pa in ca.\,240\,s (Eqn.~\ref{eqn:tau_fit}). }
\item[-] $T'$ is a dimensionless constant rescaling the independently measured structuration. \arr{Its physical meaning is very limited, beyond emphasizing the need to rescale the independently measured structuration to the proposed model, potentially linked to uncertainties in the measurements.} 
\item[-] $\varepsilon$ is the scaling parameter that relates the structure breakdown parameter to the ball radius \cite{biswas2021quantifying}. The higher the parameter, the more impactful the ball passage on destructuration of the fluid. 
\end{itemize}

\ar{Overall, t}he models of cases II and IV fit the experimental data better than the other models, showing the importance of considering the cement paste structure breakdown (or fluidization) that occurs at the passage of the ball. 
\arr{There is a clear physical motivation for including a parameter that captures paste breakdown: the passage of the ball introduces a local shear in the paste that destructures it and decreases its viscosity. As the ball shears new volumes of paste during its  descent, }
\arr{the impact of shear remains the same throughout the ball falling (during the steady-state regime studied in this work), and the current model captures this simple mode. Mathematically though, the additive decomposition of the structure parameter into structural growth and breakdown parameters is arguably unnecessary.
The improved agreement of case IV likely stems from the functional form of the time dependence of $\lambda$, and a single expression with a similar shape could conceivably apprehend the observed behavior. }
\arr{Regardless, to truly capture the complexity of the physics occurring in the cement paste, and improve its predictive ability beyond the narrow conditions explored in this work (e.g., multi-particle settling, additional orthogonal shear), a model with significantly more mathematical sophistication, coupling
build-up and breakdown through local shear rate,
would be needed \cite{varchanis2019modeling,de2018constructing}.}
\arr{In addition, such development would require testing a large number of cement pastes with an extensive range of properties. Both are} beyond the scope of this work.


\subsection{Prediction of the ball stoppage in an aging cement paste}

From Eqn. \ref{eqn:beris}, and using $\rho_{ball} - \rho_{paste}=$ 5972 $kg/m^3$, and the three radii $R$ of the balls (2, 2.5 and 3 mm), the yield stresses corresponding to ball stoppage could be calculated and are shown in Tab. \ref{tab:BallStop}. 

The static yield stress $\tau_s$ of the paste in the time window where ball stoppage occurred could also be estimated from the rheological characterization (Fig. \ref{fig:AS_static_YS}), and is shown in Tab. \ref{tab:BallStop} along with the value of dynamic yield stress $\tau_0$.

\begin{table}[h!]
\caption{Theoretically estimated yield stress of cement paste for ball stoppage (Eq. \Ref{eqn:beris}),  experimentally measured static yield stress $\tau_s$ at time of stoppage (Fig. \ref{fig:AS_static_YS} and \ref{fig:steady_velocity}),  \ar{\& experimentally measured dynamic yield stress $\tau_0$} }
\centering
\begin{tabular}{|c|c|c|c|}
\hline
\textbf{Ball radius (mm)} & {Theo. $\tau_y$ (Pa)} & {Exp. $\tau_s$ (Pa)} & {Exp. $\tau_0$ (Pa)}\\
\hline
 2.0 &  11.2 & 12.9-14.5 & 33\\
\hline
 2.5  & 14.0 & 12.9-14.5 & 33\\
 \hline
3.0 & 16.7 & 14.5-16.0 & 33\\
\hline
\end{tabular}
\label{tab:BallStop}
\end{table}

These results show that the experimentally measured value of static yield stress (measured through oscillatory rheology) seems to better match the predicted value of the critical yield stress linked to the observed ball stoppage \cite{beris1985creeping,tabuteau2007drag,ovarlez2012shear}, compared to the experimentally measured value of dynamic yield stress (measured by a flow ramp),
\arr{even when considering the experimental uncertainty on the yield stress values linked to paste volume and vane geometry.}

Looking at the paste shearing history during our experimental protocol, we tentatively explain this result. A volume element of paste on the ball trajectory has been at rest for some time ("paste age at rest", corresponding in our experiments to time durations between 120 and 420 s). During this time, the paste has flocculated, building a 3D network of particles that gets stiffer with time. When the ball approaches this fluid volume element, the fluid gets sheared. 
In that sense, the paste experiences a shear from 0 (at rest) to a finite value, matching the flow kinematics of the static yield stress measurement. 
This observation contrasts with results obtained in simple yield stress fluids, where dynamic yield stress is a good predictor for settling ball stoppage \cite{tabuteau2007drag}. 
\arr{In thixotropic fluids, the structural buildup,}
although evolving slowly during the displacement of the ball over a distance comparable to its radius, \arr{may partially explain the difference}
\cite{ferrari2021fully}. 
The multiscale nature of cement paste, where nano-sized hydrates form bridges between micron-sized particles, may also contribute to this distinctive behavior.\\
\arr{It is important to note that} the characteristic shear rate $\dot\gamma_{p}$ associated with the ball falling is smaller than the shear rate used to fluidize our paste (800 $s^{-1}$) \arr{before the dynamic yield stress measurement}.
The former scales with $v/2R$, with $v$ being the settling ball velocity and R the ball radius \cite{gueslin2006flow}, and was estimated to be at most $\dot\gamma_{p}$ $\approx$ 17 $s^{-1}$ (ball R=3 mm falling in the paste of age at rest 240 s, where $v=0.1\,m.s^{-1}$). 
\arr{It would be interesting to reevaluate the} dynamic yield stress 
after a shear at this low value of $\dot\gamma$ \arr{and evaluate if this yield stress would correctly predict ball stoppage.}
\arr{Here as well, a more sophisticated thixotropic model describing the effect of shear using a coupled structural buildup and breakdown parameter would help clarify this point. } \\
\\

Overall, these results show the ability of a rather simple thixotropic model to describe the behavior of a settling ball in a cement paste, and show the importance of properly describing the paste yielding behavior to match the flow kinematics experienced by the paste. \arr{Several limitations were discussed throughout the manuscript. These include the experimental uncertainty linked to the mixing protocol of the cement paste, which would become dominant if the model were used to predict the behavior of fresh concrete mixed in a truck from laboratory measurements on cement paste. In addition, the treatment of structural buildup and breakdown treated as independent contributions in the thixotropic model calls for the need to fitting parameters, and could not predict ball stoppage. A thixotropic model using a coupled structural buildup and breakdown parameter that captures shear effects may be able to describe the ball steady-state velocity, its slowing down, and its stoppage, in a unique model. These aspects constitute important research directions for future work.}

\section{Conclusion}

This study investigated the settling of spherical balls through cement pastes of increasing age at rest. The balls were made of metal and had three different radii (2\,mm, 2.5\,mm and 3\,mm). The results show that the average velocity of the ball decreases logarithmically with cement \ar{paste age at rest}, until the ball stops moving within the paste, a behavior that has been observed in other thixotropic materials.
\ar{In parallel, the paste was characterized using oscillatory rheology to determine its complex viscosity evolution with time at rest, the evolution of its static yield stress with time at rest, and its dynamic yield stress. Irreversible effects were measured to account for less than 15 \% of the property evolution over the timescale of the experiments, and were neglected in the analysis.}
\ar{A simple model was developed, inspired by other works on thixotropic fluids and aging yield stress fluids, that enabled us to explore several forms of the $\lambda$ parameter. Comparison of the model predictions with the experimentally measured settling-ball velocities showed improved agreement when a non-linear time-dependent structure growth parameter was employed and fluid structure breakdown was taken into account.}
Finally, \ar{the paste age at rest at which the ball stopped was recorded and compared with theory, pointing out that the measured value of static yield stress of the cement paste was a better predictor than the dynamic yield stress.}

\section{Acknowledgements}
 The work was supported by Österreichische Bautechnik Vereinigung
(ÖBV) and Österreichische Forschungsförderungsgesellschaft
(FFG) (Project number 870962). The authors thank C. Barentin and R. Biswas for fruitful discussions, and J. Kirnbauer and M. Ramizi for their help during the experimental campaign. Authors also acknowledge TU Wien Bibliothek for financial support through its Open Access Funding Program.

\arr{During the preparation of this work, the authors used Claude Sonnet 5.5, Anthropic for the coding of the model, the literature search and the editing of the manuscript. The authors reviewed and edited the output as needed and take full responsibility for the content of the published article. }

\pagebreak
\clearpage

\bibliography{settling_bibliography}

\section{Supplementary Information - 
Spherical Balls Settling Through a Quiescent
Cement Paste Measured by X-ray Tomography: 
Influence of the Paste Thixotropy}

\maketitle

\section*{SI1: Cement paste rheological behavior: reversible and irreversible evolutions}

\ar{To evaluate further the reversibility of the paste structuring behavior at rest, a longer protocol consisting of G' growth at rest (600 s), high shear (120 s at 800 rpm), and G' growth (600 s) at rest was implemented. The maximum G' obtained at the end of the first step of G' growth was used to normalize G' values. Results show that the second growth of G' (paste is 720 s older) is almost superposed to the first one until about 500\,s. In other terms, the impact of time since first contact with water (related to hydration effects) 
has significantly less impact than the time where the paste is left at rest (defined as "paste age at rest" in this work).}

\begin{figure}[!h]
\centering
\includegraphics[width=17pc]{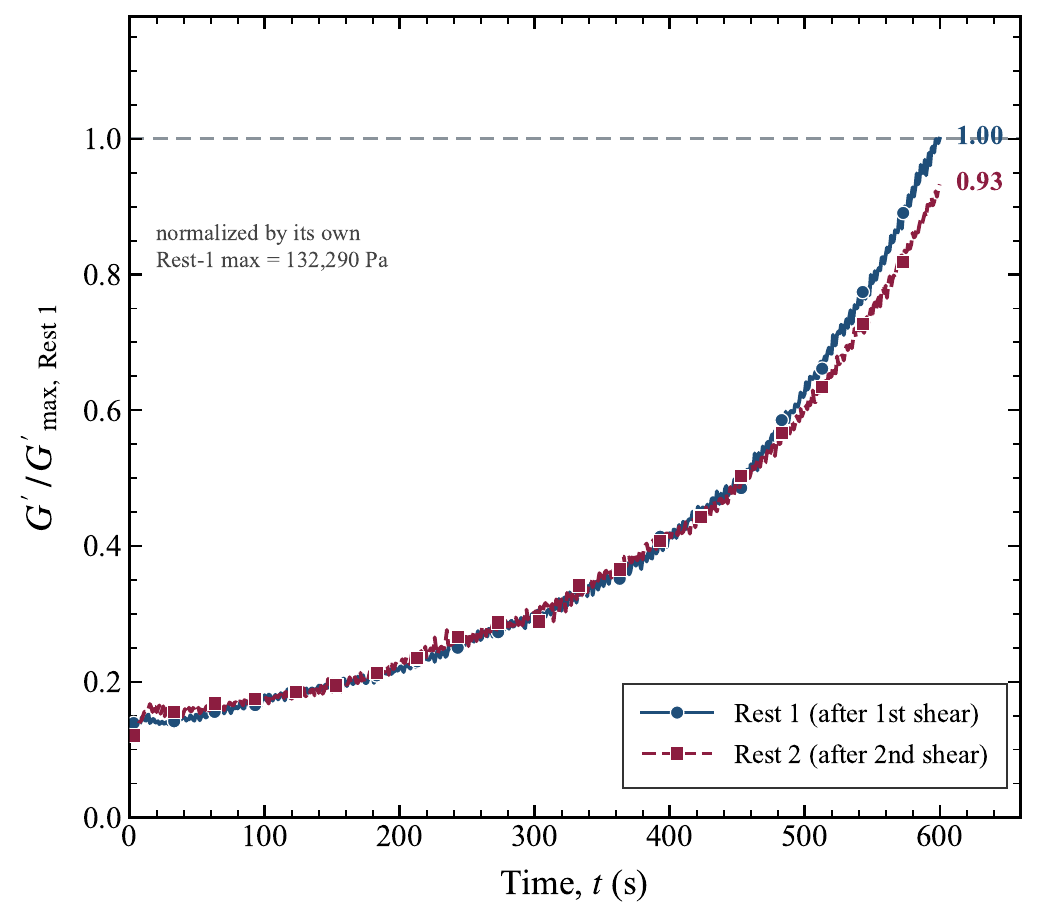}
\captionsetup{width=30pc}
\caption{G' growth from protocol consisting of G' growth at rest, high shear (2 min at 800 RPM), and G' growth at rest. The maximum G' obtained at the end of the first step of G' growth was used to normalize G' values.} 
\label{fig:G_RSR_normalized}
\end{figure}

\begin{figure}[!h]
\centering
\includegraphics[width=24pc]{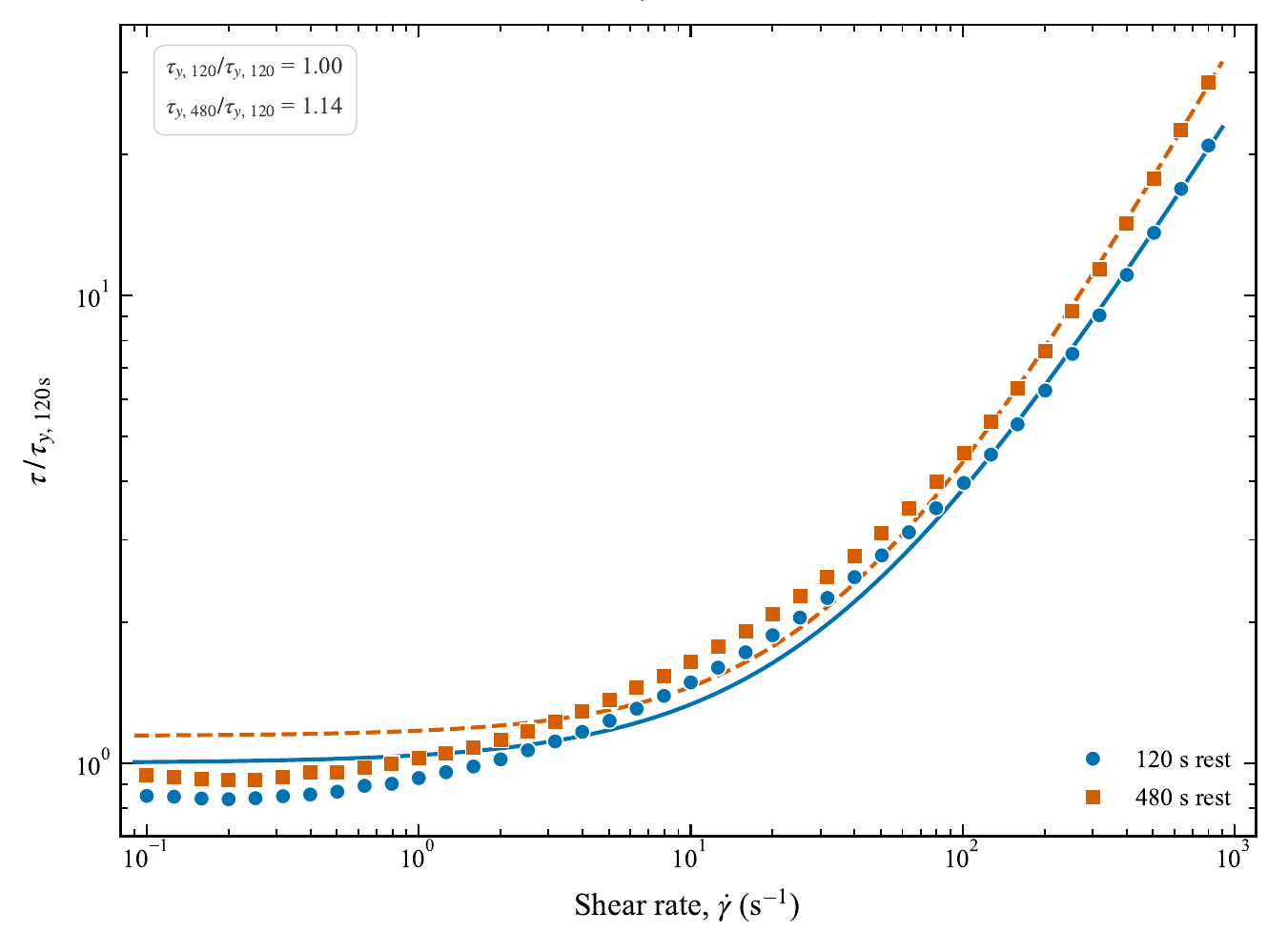}
\captionsetup{width=30pc}
\caption{Shear ramps from protocol in steps: 120 s rest, shear ramp low to high and high to low, 480 s rest,  shear ramp low to high and high to low. All stress values are normalized by the yield stress obtained from a Herschel Bulkley fit on the first ramp high to low, meant to highlight the evolution of yield stress. } 
\label{fig:DYS_SRS}
\end{figure}

\ar{The flow ramps were also performed within a longer protocol consisting of rest (duration 120 s), shear ramps low-to-high and high-to-low (duration 200 s), shear at 800 $s^{-1}$ (duration 120 s), rest (duration 480 s), shear ramps low-to-high and high-to-low (duration 200 s).
The data shown Fig. \ref{fig:DYS_SRS} present the stress normalized by the dynamic yield stress identified using a Herschel-Bulkley model on  the ramp 1 high-to-low. Results show that the stresses measured in the second ramp (where paste is 800 s older) are only slightly higher.
In particular, the dynamic yield stress identified using the second ramp is only 14 \% higher than the dynamic yield stress identified from the first ramp (800 s earlier).
}

\section*{SI2: Falling ball results}

\begin{figure}[!h]
\centering
\includegraphics[width=28pc]{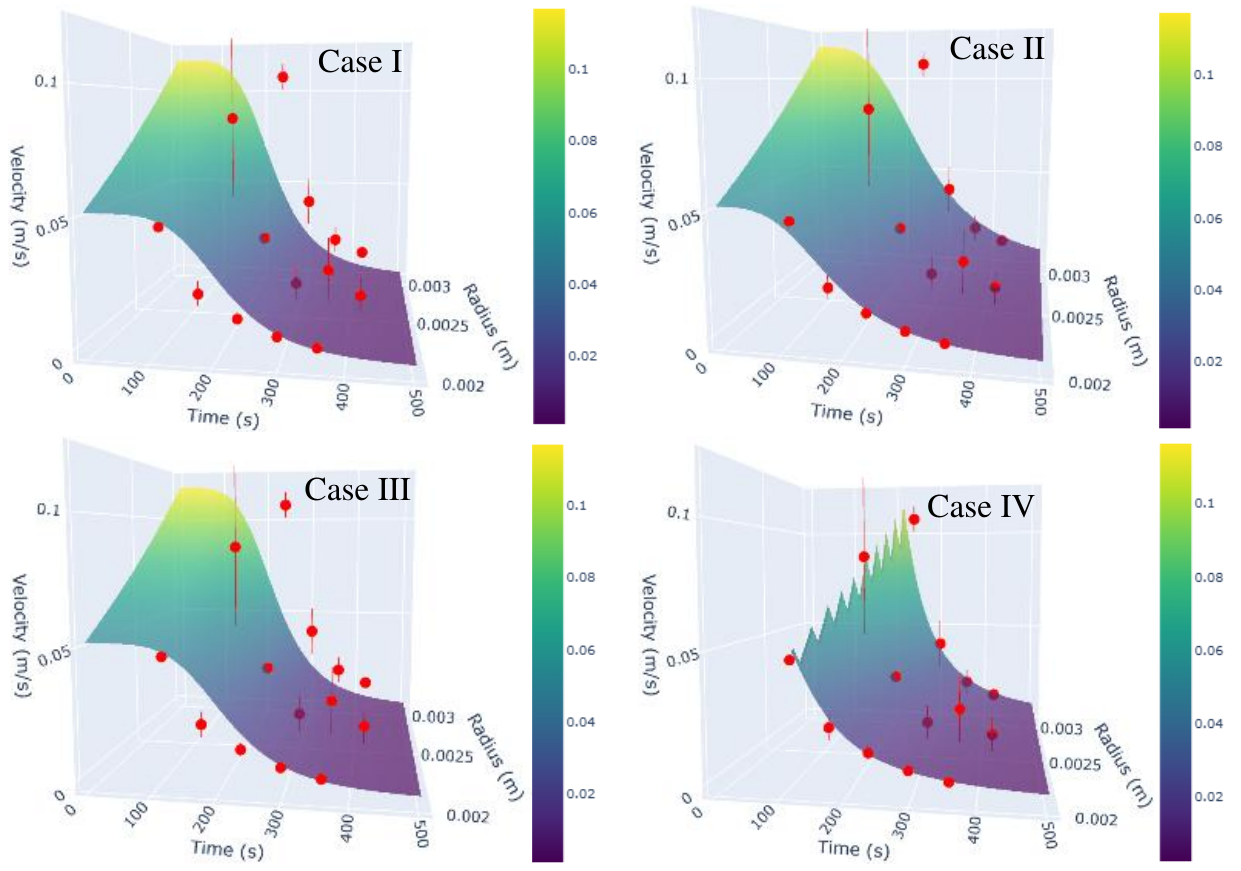}
\captionsetup{width=30pc}
\caption{Three dimensional plots to fit Cases I to IV. The bars on the right represent the colour coding of the velocity. The three axes represent, the ball velocity, time (age of paste) at start of sphere falling, and ball radius.} 
\label{fig:3D_plot}
\end{figure}

Fig. \ref{fig:3D_plot} shows the 3D fits of the ball velocity for the Cases I - IV (Eqns. 13, 15, 17 and 19 of main text). The three axes are the ball velocity, time (age of paste) and the ball radius. the input and output parameters are discussed in Section 6.1 of main text. The R-squared values in Case I to IV are 0.56, 0.70, 0.56 and 0.91 respectively. A value of R-squared closer to 1 represents a better fit. 

 \begin{figure}[!h]
\centering
\includegraphics[width=20pc]{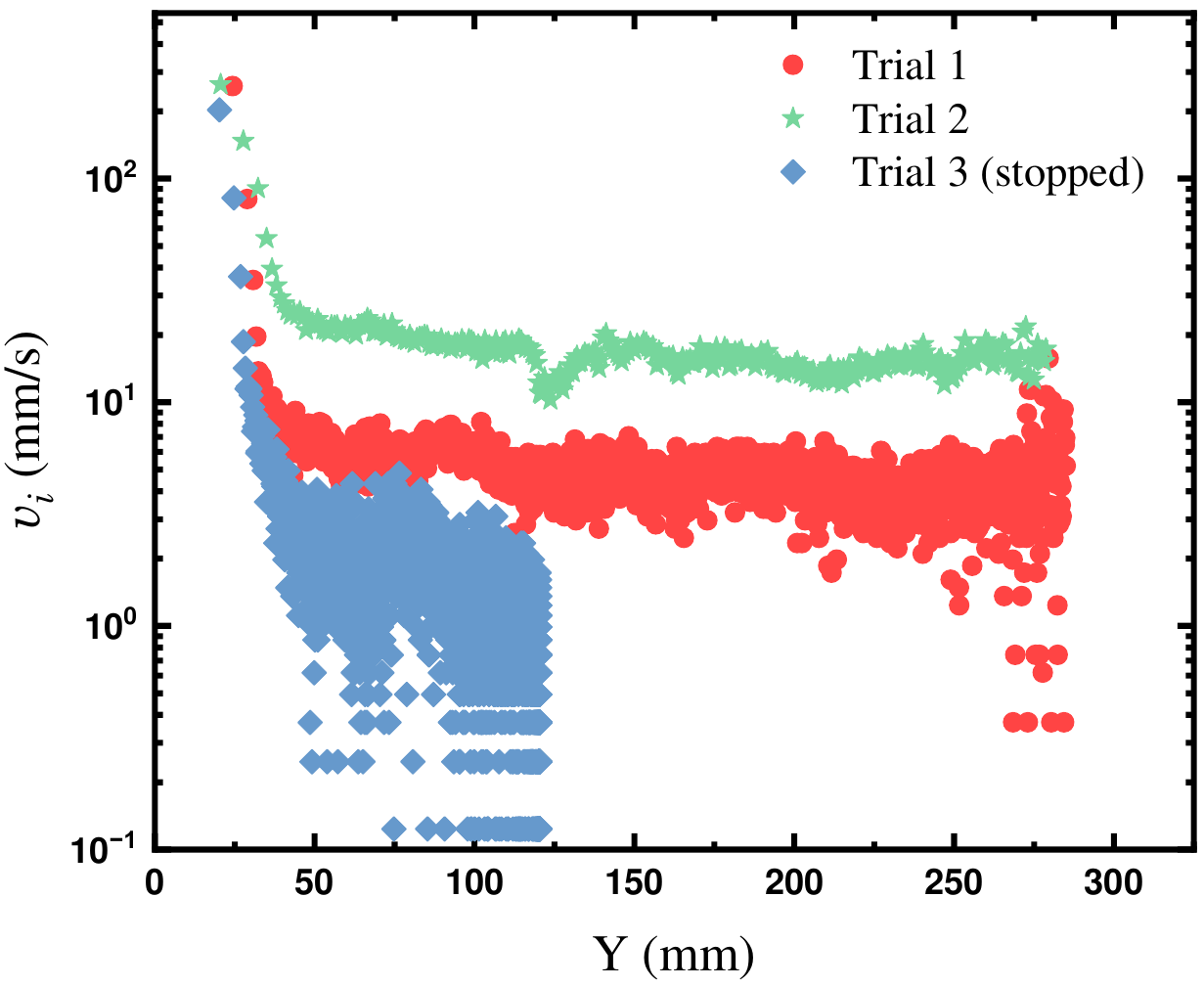}
\captionsetup{width=30pc}
\caption{Instantaneous ball velocity $v_i$ as a function of position for paste age of $t =$ 420\,s, using a 2.5\,mm ball. During trial 3 (blue diamonds), the ball stopped within the window of observation.} 
\label{fig:Slow_Down}
\end{figure}

\end{document}